\documentclass[aps, prd, amsmath, twocolumn, floats,floatfix, 10pt, superscriptaddress, nofootinbib,
showkeys]{revtex4}

\DeclareRobustCommand{\VAN}[3]{#2}
\let\VANthebibliography\thebibliography
\def\thebibliography{\DeclareRobustCommand{\VAN}[3]{##3}\VANthebibliography}

 \usepackage{lipsum}
 
\usepackage{amssymb}
\usepackage{orcidlink}
\usepackage{amsmath}
\usepackage{verbatim}
\usepackage{mathrsfs}
\usepackage{amsfonts}
\usepackage{latexsym}
\usepackage{epsfig}
\usepackage{color}
\usepackage[dvipsnames]{xcolor}
\usepackage{graphicx,subfigure}
\usepackage{units}
\usepackage{envmath}
\usepackage{natbib}
\usepackage{multirow}
\usepackage{ctable}
\usepackage{soul}
\usepackage[T1]{fontenc}
\usepackage{adjustbox}
\usepackage{float}

\begin{document}


\definecolor{orange}{rgb}{0.9,0.45,0}

\def\CovDev{D}
\def\Res{{\mathcal R}}
\def\Gammaflat{\hat \Gamma}
\def\metricflat{\hat \gamma}
\def\Dflat{\hat {\mathcal D}}
\def\part_n{\partial_\perp}

\def\Lie{\mathcal{L}}
\def\A{\mathcal{X}}
\def\Aphi{\A_{\phi}}
\def\hAphi{\hat{\A}_{\phi}}
\def\E{\mathcal{E}}
\def\Ham{\mathcal{H}}
\def\M{\mathcal{M}}
\def\R{\mathcal{R}}
\def\p{\partial}

\def\hg{\hat{\gamma}}
\def\hA{\hat{A}}
\def\hD{\hat{D}}
\def\hE{\hat{E}}
\def\hR{\hat{R}}
\def\hcA{\hat{\mathcal{A}}}
\def\hDelt{\hat{\triangle}}

\def\na{\nabla}
\def\dif{{\rm{d}}}
\def\non{\nonumber}
\newcommand{\erf}{\textrm{erf}}

\newcommand{\vdag}{(v)^\dagger}

\title{A Multi-Probe ISW Study of Dark Energy Models with Negative Energy Density: Galaxy Correlations, Lensing Bispectrum, and Planck ISW–Lensing Likelihood}

\author{Payam Ghafari
\orcidlink{0009-0008-4650-6134}}
\email{ghafaripayam09@gmail.com}
\affiliation{Department of Physics, K.N. Toosi University of Technology, P.O. Box 15875-4416, Tehran, Iran}
\affiliation{PDAT Laboratory, Department of Physics, K.N. Toosi University of Technology, P.O. Box 15875-4416, Tehran, Iran}

\author{Mahdi Najafi \orcidlink{0009-0007-1230-9880}}
\email{mahdinajafi12676@yahoo.com}
\affiliation{Dipartimento di Fisica, Univ. La Sapienza, P. le A. Moro 2, Roma, Italy}
\affiliation{PDAT Laboratory, Department of Physics, K.N. Toosi University of Technology, P.O. Box 15875-4416, Tehran, Iran}

\author{Mina Ghodsi Yengejeh \orcidlink{0000-0001-5481-9810}}
\email{mina.ghodsi@csfk.org
}
\affiliation{PDAT Laboratory, Department of Physics, K.N. Toosi University of Technology, P.O. Box 15875-4416, Tehran, Iran}
\affiliation{MTA--CSFK \emph{Lend\"ulet} ``Momentum'' Large-Scale Structure (LSS) Research Group, Konkoly Thege Mikl\'os \'ut 15-17, H-1121 Budapest, Hungary}
\affiliation{Konkoly Observatory, HUN-REN Research Centre for Astronomy and Earth Sciences, Konkoly Thege Mikl\'os \'ut 15-17, H-1121 Budapest, Hungary}
\affiliation{Institute of Physics and Astronomy, ELTE E\"otv\"os Lor\'and University, P\'azm\'any P\'eter s\'et\'any 1/A, H-1117 Budapest, Hungary}


\author{Emre \"{O}z\"{u}lker}
\email{e.ozulker@sheffield.ac.uk}
\affiliation{School of Mathematical and Physical Sciences, University of Sheffield, Hounsfield Road, Sheffield S3 7RH, United Kingdom}

\author{Eleonora Di Valentino}
\email{e.divalentino@sheffield.ac.uk}
\affiliation{School of Mathematical and Physical Sciences, University of Sheffield, Hounsfield Road, Sheffield S3 7RH, United Kingdom}

\author{Javad T. Firouzjaee}
\email{firouzjaee@kntu.ac.ir}
\affiliation{Department of Physics, K.N. Toosi University of Technology, P.O. Box 15875-4416, Tehran, Iran}
\affiliation{PDAT Laboratory, Department of Physics, K.N. Toosi University of Technology, P.O. Box 15875-4416, Tehran, Iran}
\affiliation{School of Physics, Institute for Research in Fundamental Sciences (IPM), P.O. Box 19395-5531, Tehran, Iran}


\begin{abstract}
We investigate the late-time imprints of three dark energy (DE) models, namely, the Chevallier–Polarski–Linder (CPL) parametrization, $\Lambda_{\rm s}$CDM, and an Omnipotent DE model, on cosmological observables sensitive to the time evolution of gravitational potentials. While CPL serves as a reference parameterization, the Omnipotent and $\Lambda_{\rm s}$CDM scenarios were originally proposed as possible solutions to the $H_0$ tension and are selected here because they can yield negative dark energy. These models are examined within a multi-probe framework based on the Integrated Sachs–Wolfe (ISW) effect and the lensing–ISW bispectrum. By analyzing both two- and three-point Cosmic Microwave Background (CMB) correlations, we assess how their late-time dynamics modify the growth and decay of large-scale gravitational potentials compared to the standard $\Lambda$CDM cosmology. Despite producing nearly indistinguishable CMB angular spectra at high multipoles, these models yield distinctive signatures in the low-$\ell$ ISW plateau as well as in higher-order statistics related to ISW, highlighting the power of both large-scale CMB anisotropies and higher-order CMB statistics in testing dark energy physics. Our results demonstrate that combining complementary ISW probes provides an effective way to discriminate between dark energy scenarios and will be crucial to determine whether negative or sign-switching dark energy is ultimately favored or disfavored by forthcoming data.

\end{abstract}


\maketitle

\vspace{0.8cm}
\section{Introduction}

The cosmic microwave background (CMB) encodes information about the physical conditions at the last scattering surface as well as secondary anisotropies that arise along the line of sight~\cite{White:1994sx}. The \textit{COBE} satellite first detected tiny temperature fluctuations in the CMB in the early 1990s~\cite{COBE:1992syq}, and since then these anisotropies have become one of the most powerful tools for studying the Universe. Subsequently, the \textit{Wilkinson Microwave Anisotropy Probe} (\textit{WMAP}) mission produced significantly sharper maps~\cite{WMAP:2003elm, WMAP:2003ivt, WMAP:2012fli}, and the \textit{Planck} satellite provided even more precise measurements of the temperature and polarization patterns of the CMB~\cite{Planck:2013oqw, Planck:2018nkj, Planck:2013owu, Planck:2013wtn}. In addition, ground-based surveys such as the \textit{Atacama Cosmology Telescope} (ACT)~\cite{ACT:2025fju} and the \textit{South Pole Telescope} (SPT)~\cite{SPT-3G:2025bzu} have contributed high-resolution data on smaller angular scales. Among these anisotropies, the Integrated Sachs–Wolfe (ISW) 
effect, which arises from the frequency shift of CMB photons traversing time-evolving gravitational potentials, provides a direct probe of the late-time expansion history of the Universe~\cite{Sachs:1967er}.

The ISW signal has been detected through cross-correlations between CMB temperature maps and various large-scale structure (LSS) tracers, including radio sources, infrared galaxies, and optical surveys~\cite{Cooray:2002dia, SDSS:2003lnz, Boughn:2003yz, Fosalba:2003ge, Giannantonio:2012aa, Planck:2015fcm, Velten:2015qua, Schaefer:2008qs, Granett:2015dna}. While these detections are consistent with the predictions of the $\Lambda$ Cold Dark Matter ($\Lambda$CDM) model, the ISW effect is also a sensitive probe of alternative cosmological models, as it depends on both the growth of cosmic structure and the time evolution of gravitational potentials~\cite{Hu:1994uz, Schaefer:2005up, Das:2013sca, Yengejeh:2022tpa, Y:2021ybx, Reyhani:2024cnr}.

A growing number of observations have revealed a significant tension between the values of the Hubble constant $H_0$ inferred from early- and late-universe probes~\cite{Verde:2019ivm,DiValentino:2020zio,DiValentino:2021izs,Perivolaropoulos:2021jda,Schoneberg:2021qvd,Shah:2021onj,Abdalla:2022yfr,DiValentino:2022fjm,Kamionkowski:2022pkx,Giare:2023xoc,Hu:2023jqc,Verde:2023lmm,DiValentino:2024yew,Perivolaropoulos:2024yxv,CosmoVerse:2025txj}, with local measurements consistently yielding higher values than those derived from CMB observations under the assumption of $\Lambda$CDM. The discrepancy now exceeds $7\sigma$ when the latest local distance-network measurements from the H0DN collaboration~\cite{H0DN:2025lyy} is compared with the new SPT high-$\ell$ CMB data~\cite{SPT-3G:2025bzu}. The H0DN analysis provides a new consensus value for distance-ladder determinations (see also~\cite{Freedman:2020dne,Birrer:2020tax,Anderson:2023aga,Scolnic:2023mrv,Jones:2022mvo,Anand:2021sum,Freedman:2021ahq,Uddin:2023iob,Huang:2023frr,Li:2024yoe,Pesce:2020xfe,Kourkchi:2020iyz,Schombert:2020pxm,Blakeslee:2021rqi,deJaeger:2022lit,Murakami:2023xuy,Breuval:2024lsv,Freedman:2024eph,Riess:2024vfa,Vogl:2024bum,Scolnic:2024hbh,Said:2024pwm,Boubel:2024cqw,Scolnic:2024oth,Li:2025ife,Jensen:2025aai}).
One class of possible solutions involves modifications to the dark energy (DE) sector that allow for non-standard behaviour, such as a change in the sign of the energy density or phantom crossings. In particular, models featuring negative dark energy densities at early times~\cite{DiValentino:2020naf,Adil:2023exv,Specogna:2025guo,Cheng:2025lod}, or a sign-switching cosmological constant~\cite{Akarsu:2019hmw,Akarsu:2021fol,Akarsu:2022typ,Akarsu:2023mfb,Akarsu:2024eoo,Chan:2022eyi,DiGennaro:2022ykp,Ong:2022wrs,Toda:2024ncp,Pai:2024ydi,Sabogal:2025mkp,Soriano:2025gxd,Tamayo:2025xci,Bouhmadi-Lopez:2025ggl,Bouhmadi-Lopez:2025spo}, can alter the late-time expansion rate and help to alleviate the $H_0$ tension. These models also leave potentially observable imprints on large-scale CMB anisotropies and the ISW signal, motivating an in-depth comparison with data.
This investigation is particularly timely in light of the 2025 DESI Data Release~2, which includes three years of spectroscopic Baryon Acoustic Oscillation (BAO) observations. Assuming the Chevallier–Polarski–Linder (CPL) parametrization, the DESI collaboration has reported a $2.8$--$4.2\sigma$ preference for dynamical dark energy when combining their BAO measurements~\cite{DESI:2025zgx} with \textit{Planck} CMB data~\cite{Planck:2018vyg} and several Type~Ia supernova compilations~\cite{Scolnic:2021amr,Brout:2022vxf,DES:2024hip,DES:2024jxu,DES:2024upw,Rubin:2023ovl}.\footnote{Note that the $4.2\sigma$ tension arose when the DESY5 supernova compilation was the chosen data set. This tension was reduced to $3.2\sigma$ with the recent reanalysis by the DES collaboration~\cite{DES:2025sig}, effectively leaving the tension range at $2.8$--$3.8\sigma$, where $3.8\sigma$ is obtained when the considered supernova compilation is Union3.} These results reinforce the case for non-standard dark energy models as a promising path toward resolving current cosmological tensions (see also~\cite{DESI:2024mwx,Cortes:2024lgw,Shlivko:2024llw,Luongo:2024fww,Yin:2024hba,Gialamas:2024lyw,Dinda:2024kjf,Najafi:2024qzm,Wang:2024dka,Ye:2024ywg,Tada:2024znt,Carloni:2024zpl,Chan-GyungPark:2024mlx,DESI:2024kob,Ramadan:2024kmn,Notari:2024rti,Orchard:2024bve,Hernandez-Almada:2024ost,Pourojaghi:2024tmw,Giare:2024gpk,Reboucas:2024smm,Giare:2024ocw,Chan-GyungPark:2024brx,Menci:2024hop,Li:2024qus,Li:2024hrv,Notari:2024zmi,Gao:2024ily,Fikri:2024klc,Jiang:2024xnu,Zheng:2024qzi,Gomez-Valent:2024ejh,RoyChoudhury:2024wri,Lewis:2024cqj,Wolf:2025jlc,Shajib:2025tpd,Giare:2025pzu,Chaussidon:2025npr,Kessler:2025kju,Pang:2025lvh,RoyChoudhury:2025dhe,Scherer:2025esj,Teixeira:2025czm,Specogna:2025guo,Cheng:2025lod,Cheng:2025hug,Lee:2025pzo,Ormondroyd:2025iaf,Silva:2025twg,Ishak:2025cay,Fazzari:2025lzd,Smith:2025icl}).

In this paper, we adopt a multi-probe approach to the Integrated Sachs–Wolfe (ISW) effect in order to explore the late-time imprints of non-standard dark energy models with negative energy density. We analyse both theoretical predictions and observational constraints from three complementary ISW observables: the ISW–galaxy cross-correlation, the lensing–ISW bispectrum, and the \textit{Planck} ISW–lensing likelihood. Our study focuses on three representative classes of dark energy models: (i) the CPL parametrization~\cite{Linder:2002et,Chevallier:2000qy}, which features a time-varying equation of state; (ii) the $\Lambda_{\rm s}$CDM model, which includes a sign-switching cosmological constant~\cite{Akarsu:2021fol,Akarsu:2022typ,Akarsu:2023mfb,Akarsu:2024eoo}; and (iii) the Omnipotent dark energy model~\cite{DiValentino:2020naf,Adil:2023exv,Specogna:2025guo}, which allows for both negative energy density and phantom-divide crossings. We show that these models imprint distinct signatures on both two- and three-point statistics of the CMB temperature anisotropies, particularly in the ISW–galaxy cross-correlation and the lensing–ISW bispectrum, providing a pathway to distinguish them from the standard $\Lambda$CDM scenario.
The lensing–ISW bispectrum is a secondary, late-time signal generated by the non-linear coupling between weak lensing and the time-varying ISW effect. Unlike primordial non-Gaussianity, it arises entirely from the evolution of gravitational potentials at low redshifts and is therefore highly sensitive to the dynamics of dark energy. While different models may yield similar predictions for two-point correlations, their bispectrum signatures can differ significantly, offering a complementary avenue to break degeneracies and to constrain dark energy behaviour beyond $\Lambda$CDM. Our results show that both the $\Lambda_{\rm s}$CDM and Omnipotent models can deviate markedly from $\Lambda$CDM predictions, with differences that may be detectable by current or forthcoming surveys.
Finally, we complement these theoretical investigations with real data, including the \textit{Planck} PR4 ISW–lensing likelihood~\cite{Carron:2022eum}, combined with CMB and BAO measurements, to test the viability of these models against current observations. This unified ISW framework, spanning two- and three-point correlations and connecting theoretical predictions with observational data, provides a powerful means of probing the late-time dynamics of dark energy.

This paper is structured as follows. In Section~\ref{sec:background}, we describe the homogeneous and perturbed background evolution relevant for structure formation. Section~\ref{sec:models} introduces the three dark energy models considered in this work. The data sets, including the \textit{Planck} PR4 ISW lensing likelihood, CMB, and BAO measurements, together with the numerical methods and parameter estimation procedure, are outlined in Section~\ref{sec:data}. Section~\ref{sec:ISW} presents our analysis of the ISW effect and its sensitivity to model parameters, while Section~\ref{sec:bispectrum} focuses on the lensing ISW bispectrum. In Section~\ref{sec:results}, we report the constraints from our multi probe ISW analysis based on current data. Finally, in Section~\ref{sec:conclusion}, we summarize our main findings and discuss future prospects.


\section{homogeneity and inhomogeneity}
\label{sec:background}

In this section, we outline the background evolution and perturbation formalism that apply to all models studied in this work. The expanding Universe, assumed to be homogeneous and isotropic on large scales, is described by the flat Friedmann–Lemaître–Robertson–Walker (FLRW) metric. Assuming that the cosmic medium behaves as a perfect fluid and imposing local energy–momentum conservation, the time evolution of each component is governed by
\begin{equation}
\label{eq:time}
    \rho'_i + 3\left(\frac{a'}{a}\right)(1 + w_i)\rho_i = 0 \,,
\end{equation}
where $a$ is the scale factor, and $w_i \equiv P_i / \rho_i$ denotes the barotropic equation-of-state (EoS) parameter for the $i$th component of the Universe. Here, $\rho_i$ and $P_i$ are the energy density and pressure, respectively, with $i = r, m, \text{DE}$ corresponding to radiation, matter, and dark energy, and primes denote derivatives with respect to conformal time. Equation~\eqref{eq:time} assumes that the components evolve independently, with no energy exchange between them.

The expansion dynamics of the Universe are governed by the Friedmann equation, which in a flat FLRW metric takes the form
\begin{align}
    \left(\frac{a'}{a}\right)^2 = H_0^2 a^2 \bigg[ 
    & \Omega_{r,0} a^{-4} + \Omega_{m,0} a^{-3} \nonumber \\
    & + \Omega_{\mathrm{DE},0} a^{-3} 
    \exp\left(-3 \int_1^a \frac{w_{\mathrm{DE}}(a')}{a'} \, da' \right) 
    \bigg].
\end{align}
The dimensionless density parameters are defined as $\Omega_i \equiv \rho_i / \rho_c$, where $\rho_c \equiv 3 H_0^2 / (8 \pi G)$ is the critical density of the Universe. Throughout this paper, a subscript $0$ denotes the present-day value of a quantity.

To investigate structure formation on sub-horizon scales, it is necessary to introduce perturbations to the flat FLRW metric. By adopting the synchronous gauge, all metric perturbations are contained within the spatial components, while the time–time and time–space components remain unperturbed. In this framework, the perturbed metric takes the form
\begin{equation}
    ds^2 = a^2(\tau)\left[-d\tau^2 + \left(\delta_{ij} + h_{ij}\right) dx^i dx^j\right].
\end{equation}
The metric perturbations $h_{ij}$ around the FLRW background can be decomposed into a trace part, $h \equiv h_{ii}$, and a traceless part, denoted by $h_{ij}$ for $i \neq j$. In Fourier space, the dimensionless density perturbation, $\delta_i$, defined as $\delta_i = \delta\rho_i / \rho_i$, and the velocity divergence, $\theta_i$, given by $\theta_i = \partial_j v^j_i$, evolve according to the following equations~\cite{Ma:1995ey}:
\begin{align}
\delta'_{\mathrm{DE}} &= - (1 + w_{\mathrm{DE}}) \left( \theta_{\mathrm{DE}} + \frac{h'}{2} \right)
- 3 \left( \frac{a'}{a} \right) (c^2_{s\,\mathrm{DE}} - w_{\mathrm{DE}}) \nonumber \\
&\quad \times \left[ \delta_{\mathrm{DE}} + 3 \left( \frac{a'}{a} \right)
\frac{(1 + w_{\mathrm{DE}}) \theta_{\mathrm{DE}}}{k^2} \right]
 \nonumber \\
&\quad - 3 \left( \frac{a'}{a} \right)
\frac{w'_{\mathrm{DE}} \theta_{\mathrm{DE}}}{k^2}, \label{eq:delta_de}\\
\theta'_{\mathrm{DE}} &= - \frac{a'}{a} (1 - 3 c^2_{s\,\mathrm{DE}})\, \theta_{\mathrm{DE}}
+ \frac{c^2_{s\,\mathrm{DE}}}{1 + w_{\mathrm{DE}}}\, k^2 \delta_{\mathrm{DE}}, \label{eq:theta_de}\\
\delta'_c &= - \left( \theta_c + \frac{h'}{2} \right), \label{eq:delta_c}\\
\theta'_c &= - \frac{a'}{a} \theta_c. \label{eq:theta_c}
\end{align}
The squared sound speed of the dark energy in its rest frame is given by $c_{s\rm DE}^2 \equiv \big( \delta P_{\rm DE} / \delta \rho_{\rm DE} \big)_{\rm rest frame}$. Under the assumption of a barotropic fluid, where $P_{\rm DE} = P_{\rm DE} (\rho_{\rm DE})$, the adiabatic sound speed satisfies $c_{\rm s}^2 = c_{\rm a}^2 = w_{\rm DE}$. In case of a negative EoS where $w_{\rm DE} < 0$, this results in $c^2_{\rm DE} < 0$, leading to instabilities in the dark energy fluid. To avoid this issue, we adopt $c^2_{\rm DE} = 1$ throughout our analysis.

Building on this cosmological framework, we now investigate three alternative dark energy models as potential solutions to the anomalies observed in modern cosmology.

\section{Dark Energy Models}\label{sec:models}

In this section, we examine a set of dynamical dark energy models and their implications for key cosmological observables. Among the various parameterizations proposed in the literature, we focus on two models that have only recently been introduced and that not only allow for an evolving dark energy component but have also been proposed as possible solutions to the $H_0$ tension. Their predictions are compared with those of the well-established CPL model, with particular attention to their effects on the matter power spectrum, the evolution of the gravitational potential, the ISW effect, and the lensing–ISW bispectrum.

\subsection{CPL model}

As a baseline for comparison of Dynamical DEs, we consider the CPL model~\cite{Linder:2002et,Chevallier:2000qy}, a widely used two-parameter description of the dark energy EoS that is linear in  the scale factor $a$:
\begin{equation}
    w_{\mathrm{DE}}(a) = w_0 + w_a (1 - a),
\end{equation}
where $w_0$ corresponds to the present-day value of the EoS and $w_a$ describe its evolution. Despite its simplicity, this parametrization can accurately capture the phenomenology of a wide variety of theoretical models of late time dark energy~\cite{dePutter:2008wt} and it serves as a standard benchmark for assessing the performance of more complex dark energy scenarios.

\subsection{$\Lambda_{\mathrm{s}}$CDM model}

The $\Lambda_{\mathrm{s}}$CDM model was first introduced in Ref.~\cite{Akarsu:2021fol} as a phenomenological extension of the standard $\Lambda$CDM framework, motivated by the hypothesis of a spontaneous anti–de Sitter to de Sitter transition in the Universe. This transition involves a sign change in the cosmological constant occurring around redshift $z \sim 2$.

The simplest realization of the $\Lambda_{\mathrm{s}}$CDM model is obtained by replacing the constant $\Lambda$ term of standard $\Lambda$CDM with a sign-switching counterpart, $\Lambda_{\mathrm{s}}$, which changes sign at a characteristic redshift $z^{\dagger}$—the model’s only additional free parameter. The present-day value of the dark energy density is denoted by $\Lambda_{\mathrm{s0}}$, and the evolution of $\Lambda_{\mathrm{s}}$ is given by
\begin{equation}
    \Lambda_{\mathrm{s}} = \Lambda_{\mathrm{s0}}\, \mathrm{sgn}[z^{\dagger} - z],
\end{equation}
where the signum function, $\mathrm{sgn}$, models a sharp transition in the dark energy density from negative to positive at $z = z^{\dagger}$. 

This formulation allows for a shift in the constant and has been explored as a possible mechanism to alleviate the $H_0$ tension~\cite{Akarsu:2021fol,Akarsu:2022typ,Akarsu:2023mfb,Akarsu:2024eoo}, with the simple discontinuous step function usually understood as an effective phenomenological description of an underlying model in which the transition is smooth but very rapid; see Refs.~\cite{Anchordoqui:2023woo,Anchordoqui:2024gfa,Anchordoqui:2024dqc,Akarsu:2024qsi,Souza:2024qwd,Akarsu:2025gwi} for theoretical frameworks that realize the $\Lambda_{\rm s}$CDM phenomenology.

\subsection{Omnipotent Dark Dnergy}

We now turn to a phenomenologically flexible class of scenarios called Omnipotent DE models, introduced in Refs.~\cite{DiValentino:2020naf,Adil:2023exv,Specogna:2025guo} as a possible solution to the $H_0$ tension. These models allow the dark energy density to become negative, evolve non-monotonically, and oscillate with an EoS that may include singularities and crossings of the phantom divide ($w_{\mathrm{DE}} = -1$). This flexibility enables the Omnipotent models to capture a wide range of late-time expansion histories.

Although such behaviour may appear to violate standard energy conditions, Omnipotent DE is treated as an effective source term in the Friedmann equations rather than as a fundamental component of the energy–momentum tensor, thereby preserving theoretical consistency. In contrast to phantom models, which assume $w_{\mathrm{DE}} < -1$ with $\rho_{\mathrm{DE}} > 0$, the Omnipotent framework also accommodates scenarios where $\rho_{\mathrm{DE}} < 0$ and $w_{\mathrm{DE}} > -1$ for which the energy density still increases with the expansion despite $w_{\mathrm{DE}} > -1$. This distinction follows directly from the continuity equation,
\begin{equation}\label{de}
    \frac{d \rho_{\mathrm{DE}}(z)}{d z} = 3\,\frac{1 + w_{\mathrm{DE}}(z)}{1 + z}\, \rho_{\mathrm{DE}}(z),
\end{equation}
which allows both positive and negative energy densities depending on the functional form of $w_{\mathrm{DE}}(z)$. A complete classification of the six possible combinations of sign and EoS behaviour is provided in Table~\ref{tab:omni}. The Omnipotent model stands out by encompassing all of these regimes within a single parameter space.

Following Ref.~\cite{DiValentino:2020naf}, such a dark energy density can be modeled to exhibit an extremum at scale factor $a_m$, taking the form
\begin{equation}\label{Om1}
    \rho_{\mathrm{DE}}(a) = \rho_{\mathrm{DE0}} \,
    \frac{1 + \alpha (a - a_m)^2 + \beta (a - a_m)^3}
         {1 + \alpha (1 - a_m)^2 + \beta (1 - a_m)^3},
\end{equation}
where $\alpha$, $\beta$, and $a_m$ are free parameters. The corresponding equation of state is then given by
\begin{equation}
    w_{\mathrm{DE}}(a) = -1 -
    \frac{a \left[2\alpha (a - a_m) + 3\beta (a - a_m)^2\right]}
         {3\left[1 + \alpha (a - a_m)^2 + \beta (a - a_m)^3\right]}.
\end{equation}
This parametrization belonging to the class of Omnipotent DE models is simply dubbed the Omnipotent DE hereafter.

We next investigate the impact of these models on cosmological observables using current data.

\begin{table}[t]
\caption{
Overview of the six possible DE regimes, classified by the sign of $\rho_{\mathrm{DE}}$ and the EoS parameter $w_{\mathrm{DE}}$ relative to the phantom divide ($w_{\mathrm{DE}} = -1$). The table summarizes the corresponding scaling behaviour with redshift $z$ and scale factor $a$, and assigns descriptive labels to each regime. Adapted from Ref.~\cite{Adil:2023exv}.
}
\begin{center}
\resizebox{0.48\textwidth}{!}{
\begin{tabular}{c|c|c|c|c}
    \hline \hline
    \textbf{Density} & \textbf{EoS} & \textbf{Scaling in $z$} & \textbf{Scaling in $a$} & \textbf{Name} \\
    \hline\hline
    \multirow{3}{*}{$\rho > 0$}
    & $w > -1$ & $\frac{d\rho}{dz} > 0$ & $\frac{d\rho}{da} < 0$ & p-quintessence \\
    & $w = -1$ & $\frac{d\rho}{dz} = 0$ & $\frac{d\rho}{da} = 0$ & positive-$\Lambda$ \\
    & $w < -1$ & $\frac{d\rho}{dz} < 0$ & $\frac{d\rho}{da} > 0$ & p-phantom \\
    \hline
    \multirow{3}{*}{$\rho < 0$}
    & $w > -1$ & $\frac{d\rho}{dz} < 0$ & $\frac{d\rho}{da} > 0$ & n-quintessence \\
    & $w = -1$ & $\frac{d\rho}{dz} = 0$ & $\frac{d\rho}{da} = 0$ & negative-$\Lambda$ \\
    & $w < -1$ & $\frac{d\rho}{dz} > 0$ & $\frac{d\rho}{da} < 0$ & n-phantom \\
    \hline \hline
\end{tabular}
}
\end{center}
\label{tab:omni}
\end{table}
\begin{table}[]
    \centering
    \begin{tabular}{c|c|c|c|c} %
    \hline \hline 
 $Parameter$&$\Lambda$CDM &      $\Lambda_{\rm s}$CDM& $CPL$&$Omnipotent$\\ \hline   
 $\Omega_bh^2$&$0.023$& $0.023$& $0.023$&$0.023$\\  
 $\Omega_ch^2$&$0.120$& $0.119$& $0.119$&$0.118$\\   
 $H_0$& $67.530$& $72.750$& $85.540$&$97.690$\\   
         $\tau$&  $0.047$& $0.051$& $0.046$&$0.053$\\   
         $100 \theta_{MC}$& 
     $1.041$& $1.041$& $1.041$&$1.041$\\  
 $10^{9}A_S$&$2.080$& $2.081$& $2.063$&$2.079$\\  
 $n_s$&$0.964$& $0.969$& $0.969$&$0.972$\\  \hline 
 $w_0$&$-1.000$ & $-1.000$& $-1.066$&$-1.000$\\ 
 $w_a$&$-$ & $-$& $-2.256$&$-$\\ \hline
 $\alpha$&$-$ & $-$& $-$&$1.202$\\  
 $\beta$&$-$ & $-$& $-$&$26.629$\\  
 $a_m$&$-$ & $-$& $-$&$0.591$\\   \hline
 $z^{\dagger}$&$-$ & $1.706$& $-$&$-$\\ \hline \hline \end{tabular}
    \caption{
Best-fit cosmological parameters for the considered models, derived using CMB data.
}
    \label{tab:CMB}
\end{table}

\begin{table}[]
    \centering
    \begin{tabular}{c|c|c|c|c} %
    \hline \hline
 $Parameter$&$\Lambda$CDM &      $\Lambda_{\rm s}$CDM& $CPL$&$Omnipotent$\\ \hline\hline   
 $\Omega_bh^2$&$0.023$& $0.022$& $0.022$&$0.023$\\   
 $\Omega_ch^2$&$0.119$& $0.121$& $0.120$&$0.119$\\  
 $H_0$& $67.939$& $68.353$& $64.955$&$74.476$\\  
         $\tau$&  $0.060$& $0.052$& $0.051$&$0.053$\\   
         $100 \theta_{MC}$& 
     $1.041$& $1.041$& $1.041$&$1.041$\\  
 $10^{9}A_S$&$2.107$& $2.092$& $2.087$&$2.085$\\  
 $n_s$&$0.970$& $0.964$& $0.967$&$0.967$\\ \hline  
 $w_0$&$-1.000$ & $-1.000$& $-0.550$&$-1.000$\\  
 $w_a$&$-$ & $-$& $-1.436$&$-$\\ \hline  
 $\alpha$&$-$ & $-$& $-$&$13.119$\\ 
 $\beta$&$-$ & $-$& $-$&$28.698$\\   
 $a_m$&$-$ & $-$& $-$&$0.847$\\ \hline  
 $z^{\dagger}$&$-$ & $2.661$& $-$&$-$\\ \hline \hline \end{tabular}
    \caption{
Best-fit cosmological parameters for the considered models, derived using CMB and BAO data.
}
    \label{tab:CMB+BAO}
\end{table}

\begin{table}[]
    \centering
    \begin{tabular}{c|c|c|c|c} %
    \hline \hline 
 $Parameter$&$\Lambda$CDM &      $\Lambda_{\rm s}$CDM& $CPL$&$Omnipotent$\\ \hline\hline   
 $\Omega_bh^2$&$0.023$& $0.022$& $0.023$&$0.023$\\   
 $\Omega_ch^2$&$0.119$& $0.120$& $0.118$&$0.120$\\  
 $H_0$& $68.027$& $69.816$& $77.650$&$92.970$\\  
         $\tau$&  $0.057$& $0.051$& $0.058$&$0.048$\\ 
         $100 \theta_{MC}$& 
     $1.041$& $1.041$& $1.041$&$1.041$\\  
 $10^{9}A_S$&$2.103$& $2.085$& $2.106$&$2.074$\\  
 $n_s$&$0.969$& $0.965$& $0.971$&$0.965$\\ \hline  
 $w_0$&$-1.000$ & $-1.000$& $-1.023$&$-1.000$\\  
 $w_a$&$-$ & $-$& $-1.147$&$-$\\ \hline  
 $\alpha$&$-$ & $-$& $-$&$7.582$\\   
 $\beta$&$-$ & $-$& $-$&$6.862$\\   
 $a_m$&$-$ & $-$& $-$&$0.117$\\ \hline  
 $z^{\dagger}$&$-$ & $2.192$& $-$&$-$\\ \hline \hline \end{tabular}
    \caption{
Best-fit cosmological parameters for the considered models, derived using CMB and ISW data.
}
    \label{tab:CMB+ISW}
\end{table}


\section{DATA and METHODOLOGY}\label{sec:data}

In this section, we describe the cosmological datasets and computational tools used to constrain the parameters of the dark energy models through Bayesian inference and to determine the corresponding best-fit cosmologies.

\begin{itemize}

    \item \textbf{Computational tools}: 
    Theoretical predictions for the CMB power spectra and related cosmological observables are computed using the Boltzmann solver \texttt{CAMB}~\cite{Lewis:1999bs,Howlett:2012mh}. 
    Parameter estimation is performed with the Markov Chain Monte Carlo (MCMC) sampler implemented in \texttt{Cobaya}\footnote{\url{https://ascl.net/1910.019}}~\cite{Torrado:2020dgo}, which enables efficient exploration of the parameter space and provides statistically robust posterior distributions. 
    Up to Sec.~\ref{sec:results}, we adopt the best-fit cosmologies obtained from \texttt{Cobaya}, as reported in Tables~\ref{tab:CMB}--\ref{tab:CMB+ISW}. 
    In Sec.~\ref{sec:results}, we further explore the posterior distributions of each model using the priors listed in Table~\ref{tab:prior}. 
    All MCMC chains are required to satisfy the Gelman–Rubin convergence criterion~\cite{Gelman:1992zz}, $R - 1 < 0.03$.

    \item \textbf{Statistical analysis}: 
    We assess the statistical performance of each dark energy model relative to $\Lambda$CDM using two complementary approaches:
    \begin{enumerate}
        \item \textbf{Minimum chi-square}: 
        We compute the difference $\Delta \chi^2 = \chi^2_{\mathrm{min}}(\mathrm{Model}) - \chi^2_{\mathrm{min}}(\Lambda \mathrm{CDM})$ to determine whether a given model provides a better fit to the data. 
        Negative (positive) values of $\Delta \chi^2$ indicate that the model is preferred (disfavoured) relative to $\Lambda$CDM.

        \item \textbf{Bayesian evidence}: 
        We evaluate the relative log-Bayesian evidence $\ln \mathcal{B}_{ij}$ using a modified version of the \texttt{MCEvidence}\footnote{\href{https://github.com/williamgiare/wgcosmo/tree/main}{github.com/williamgiare/wgcosmo/tree/main}}$^,$\footnote{\href{https://github.com/yabebalFantaye/MCEvidence}{github.com/yabebalFantaye/MCEvidence}} package~\cite{Heavens:2017hkr,Heavens:2017afc}, fully compatible with \texttt{Cobaya}. 
        The relative log-evidence $\Delta \ln B_{ij} = \ln B_i - \ln B_j$ quantifies the statistical preference for model $i$ over a reference model $j$ (in our case, $\Lambda$CDM). 
        Negative values of $\ln \mathcal{B}_{ij}$ indicate a preference for the extended model. 
        The interpretation of $\ln \mathcal{B}_{ij}$ follows the scale summarized in Table~\ref{tab:BE}.
    \end{enumerate}

    \item \textbf{CMB}: 
    We employ temperature and polarization data from the \textit{Planck} 2018 legacy release, including high-$\ell$ Plik TT spectra ($30 \leq \ell \lesssim 2500$), TE and EE spectra ($30 \leq \ell \lesssim 2000$), as well as low-$\ell$ TT-only ($2 \leq \ell \leq 29$) and EE-only ($2 \leq \ell \leq 29$) likelihoods. 
    We also include the CMB lensing from \textit{Planck PR4} maps as a complementary probe of the late-time gravitational potential~\cite{Planck:2018vyg,Planck:2019nip,Carron:2022eyg}.

    \item \textbf{BAO}: 
    We use Baryon Acoustic Oscillation measurements from multiple spectroscopic surveys, including SDSS Data Release 7 (DR7)~\cite{Ross:2014qpa} Main Galaxy Sample (MGS) and Data Release 16 (DR16) measurements from several tracers—Luminous Red Galaxies (LRGs), Emission Line Galaxies (ELGs), Quasars (QSOs), and Lyman-$\alpha$ (Ly$\alpha$) forests~\cite{eBOSS:2020yzd}. 
    We also include the BAO constraint from the 6dF Galaxy Survey (6dFGS)~\cite{Beutler:2011hx}. 

    \item \textbf{ISW}: 
    We include the observational ISW likelihood built from the cross-correlation between \textit{Planck} temperature maps and the CMB lensing convergence field~\cite{Carron:2022eum}. 
    This measurement isolates the late-time correlation between the ISW effect and weak gravitational lensing, providing a direct probe of the time variation of gravitational potentials. 
    It offers an independent constraint on dark energy models that alter the late-time dynamics of the Universe beyond the $\Lambda$CDM prediction.
    
\end{itemize}

\begin{table}[]
    \centering
    \begin{tabular}{c|c}\hline \hline
 $Parameter$&$Prior$\\\hline\hline
 $\Omega_bh^2$&$[0.005, 1]$\\
 $\Omega_ch^2$&$[0.01 , 0.99]$\\
         $\tau$&  $[0.01 , 0.8]$\\
         $100 \theta_{MC}$& 
     $[0.5 , 10]$\\
 $log(10^{10}A_S)$&$[1.63 , 3.91]$\\
 $n_s$&$[0.8 , 1.2]$\\\hline
 $w_0$&$[-3 , 1]$\\
 $w_a$&$[-3 , 2]$\\\hline
 $\alpha$&$[0 , 30]$\\
 $\beta$&$[0 , 30]$\\
 $a_m$&$[0 , 1]$\\\hline
 $z_{\dagger}$&$[1 , 3]$\\\hline \hline \end{tabular}
    \caption{flat prior distribution for cosmological parameters}
    \label{tab:prior}
\end{table}
\begin{table}
\centering
\renewcommand{\arraystretch}{1.5}
\begin{tabular}{c @{\hspace{1cm}}|@{\hspace{1cm}} c}
\hline \hline
$\ln \mathcal{B}_{ij}$ & \textbf{interpretation} \\
\hline \hline
$0 \leq | \ln \mathcal{B}_{ij}|  < 1$ & Inconclusive \\
$1 \leq | \ln \mathcal{B}_{ij}|  < 2.5$ & Weak \\
$2.5 \leq | \ln \mathcal{B}_{ij}|  < 5$ & Moderate \\
$5 \leq | \ln \mathcal{B}_{ij}|  < 10$ & Strong \\
$| \ln \mathcal{B}_{ij} | \geq 10$ & Very strong \\

\hline \hline
\end{tabular}
\caption{Description of the interpretation of the logarithm of the Bayes factor, $\ln \mathcal{B}_{ij}$. This representation makes it straightforward to identify which models are statistically favored or disfavored relative to each other.}
\label{tab:BE}
\end{table}

\section{Cosmological Signatures of the Dark Energy Models}\label{sec:Cosmological_sig}

In this section, we present the effects of our dark energy models on both primary and large-scale secondary cosmological observables. 
For each dataset combination (CMB only, CMB+BAO, and CMB+ISW), we use the corresponding best-fit cosmological parameters obtained from the dedicated likelihood analyses described in Sec.~\ref{sec:data} and listed in Tables~\ref{tab:CMB}-\ref{tab:CMB+ISW}. 
These best fits are not used simultaneously; instead, the appropriate set is selected depending on the observable under investigation. 
The resulting predictions for the CMB temperature anisotropy spectra and the matter power spectra are then compared with those of the standard $\Lambda$CDM cosmology. 
We further examine how variations in the free parameters of each model affect the low-multipole CMB temperature spectrum, emphasizing the distinctive signatures that differentiate them from $\Lambda$CDM.

\subsection{CMB temperature anisotropy power spectrum} \label{sec:TT}

\begin{figure}[htbp]
    \centering
    \includegraphics[width=1\linewidth]{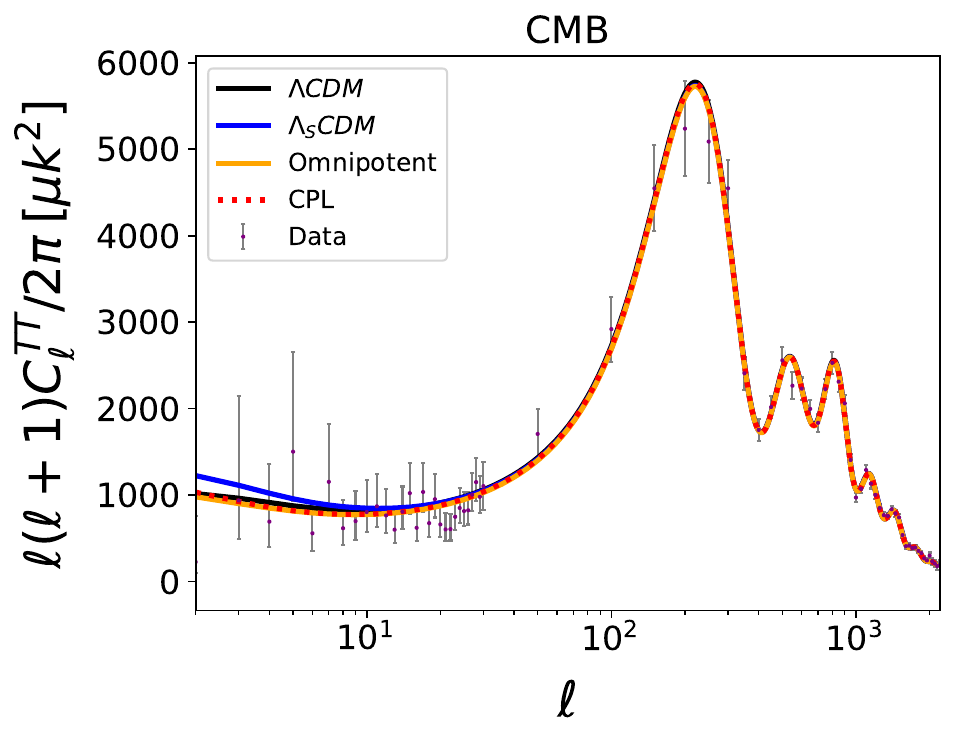}
    \caption{
CMB temperature–temperature angular power spectrum ($C_\ell^{TT}$) for the $\Lambda$CDM (black solid), CPL (red dotted), $\Lambda_{\mathrm{s}}$CDM (blue solid), and Omnipotent DE (orange solid) models, computed using the CMB-only best-fit parameters. 
\textit{Planck} 2018 data points with error bars are shown for comparison.
}
    \label{fig:TT_Power}
\end{figure}

Fig.~\ref{fig:TT_Power} shows the CMB temperature–temperature angular power spectrum, $C_\ell^{TT}$, for $\Lambda$CDM and the three dark energy models, plotted together with the \textit{Planck} 2018 data and associated error bars. 
All models accurately reproduce the acoustic peak structure up to the sixth peak, displaying an excellent level of agreement with the data at intermediate and high multipoles. 
As expected, the impact of the dark energy sector becomes relevant only at large angular scales ($\ell \lesssim 100$), where the late-time ISW contribution is significant. 
However, these differences are confined to the cosmic-variance–limited regime, making the models effectively indistinguishable from a CMB temperature point of view. 
The overall consistency demonstrates the well known fact that late-time modifications of the expansion history are not excluded or strongly preferred over $\Lambda$ by current temperature anisotropy data alone.

\subsection{Parameter sensitivity and deviations from $\Lambda$CDM}
\label{ssec:parameter-sensitivity}

To isolate the imprint of each model's free parameters on the large-scale CMB anisotropies, we examine the ratio of their temperature–temperature spectra to that of $\Lambda$CDM, focusing on the multipole range $2 \leq \ell \leq 100$. 
For this analysis, we adopt the CMB+BAO best-fit cosmological parameters of each model and vary only the model-specific free parameters, while keeping all other parameters fixed at their best-fit values. 

\begin{figure}[htbp]
    \centering
    \includegraphics[width=0.95\linewidth]{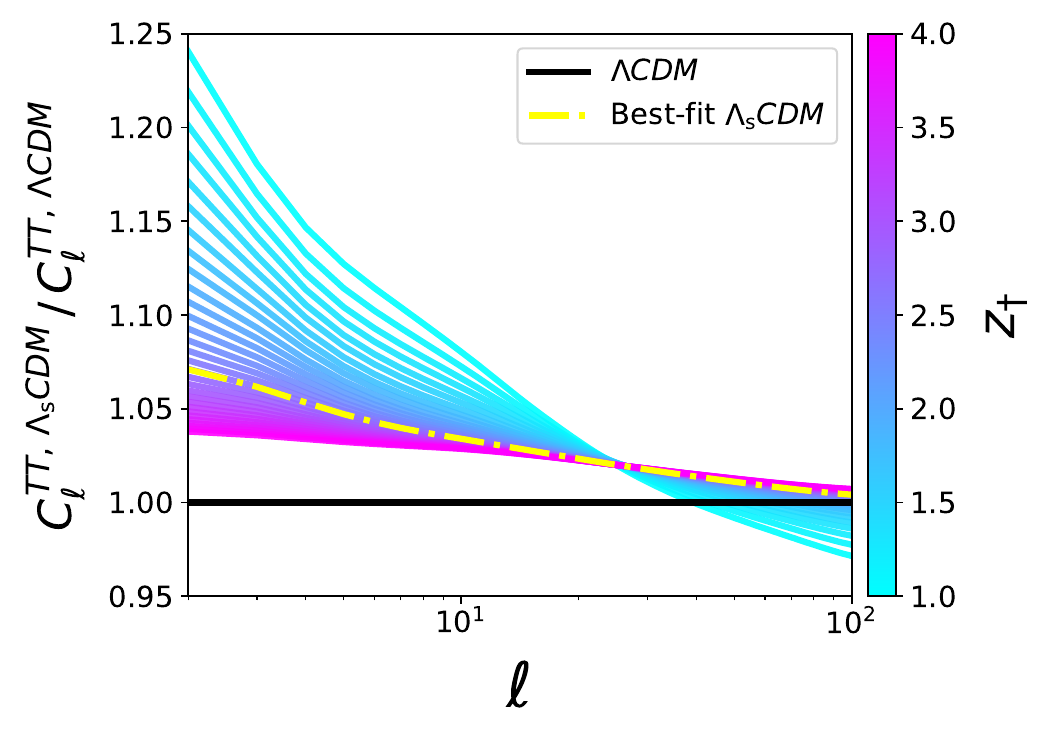}
    \caption{
Ratio of the CMB temperature power spectrum, $(C_\ell^{TT,\Lambda_{\mathrm{s}}\mathrm{CDM}} / C_\ell^{TT,\Lambda\mathrm{CDM}})$, for different values of the transition redshift $z^{\dagger}$ in the $\Lambda_{\mathrm{s}}$CDM model. 
The highlighted curve corresponds to the best-fit parameter values.
}
    \label{fig:CTT_zdag_combined}
\end{figure}

For the $\Lambda_{\rm s}$CDM model, we vary the transition redshift $z^{\dagger}$ across the range $1.0 \leq z^{\dagger} \leq 4.0$. 
The corresponding ratio plot (Fig.~\ref{fig:CTT_zdag_combined}) shows that smaller values of $z^{\dagger}$ lead to a marked amplification of the low-$\ell$ temperature power, enhancing the deviation from $\Lambda$CDM at the largest angular scales. 
Conversely, larger $z^{\dagger}$ values progressively recover the $\Lambda$CDM spectrum. 
As expected, in the limit $z^{\dagger} \to \infty$, the sign-switching cosmological constant remains positive throughout cosmic history, and the model effectively reduces to $\Lambda$CDM. 
In the figure, the $\Lambda$CDM reference is indicated by a solid black horizontal line at unity, and the best-fit $\Lambda_{\rm s}$CDM curve is highlighted with a dashed line. 
The curves are colour-coded to illustrate the parameter progression across the multipole range.

\begin{figure}[H]
    \centering
    
    \includegraphics[width=0.95\linewidth]{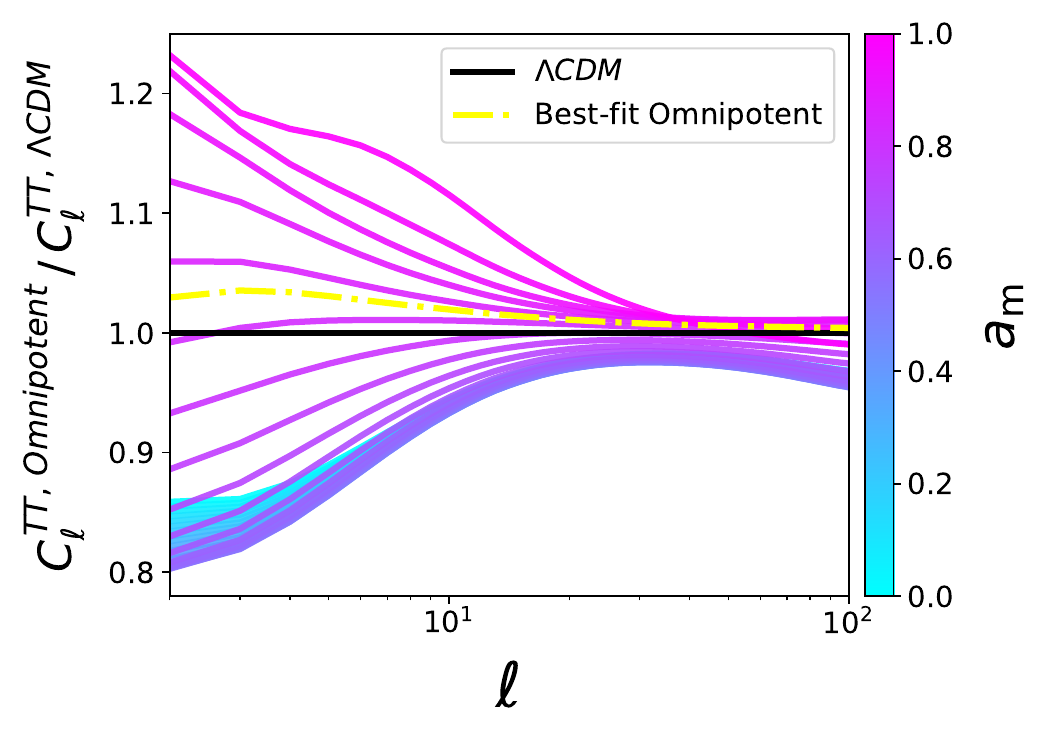}
    \includegraphics[width=0.95\linewidth]{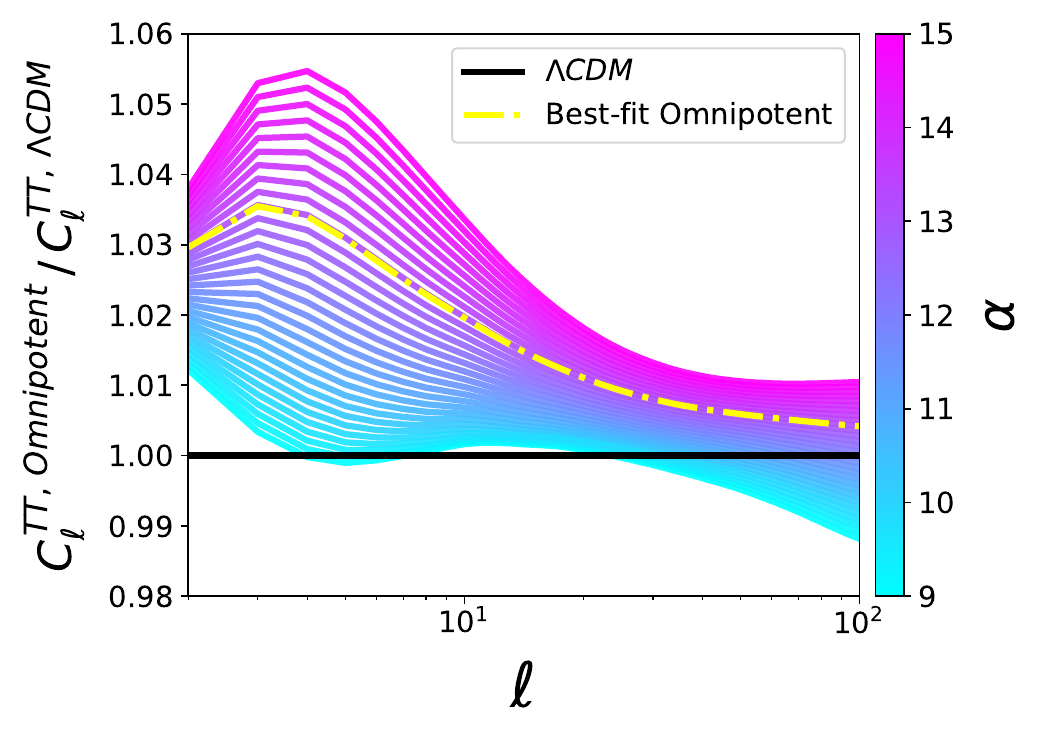}
    \includegraphics[width=0.95\linewidth]{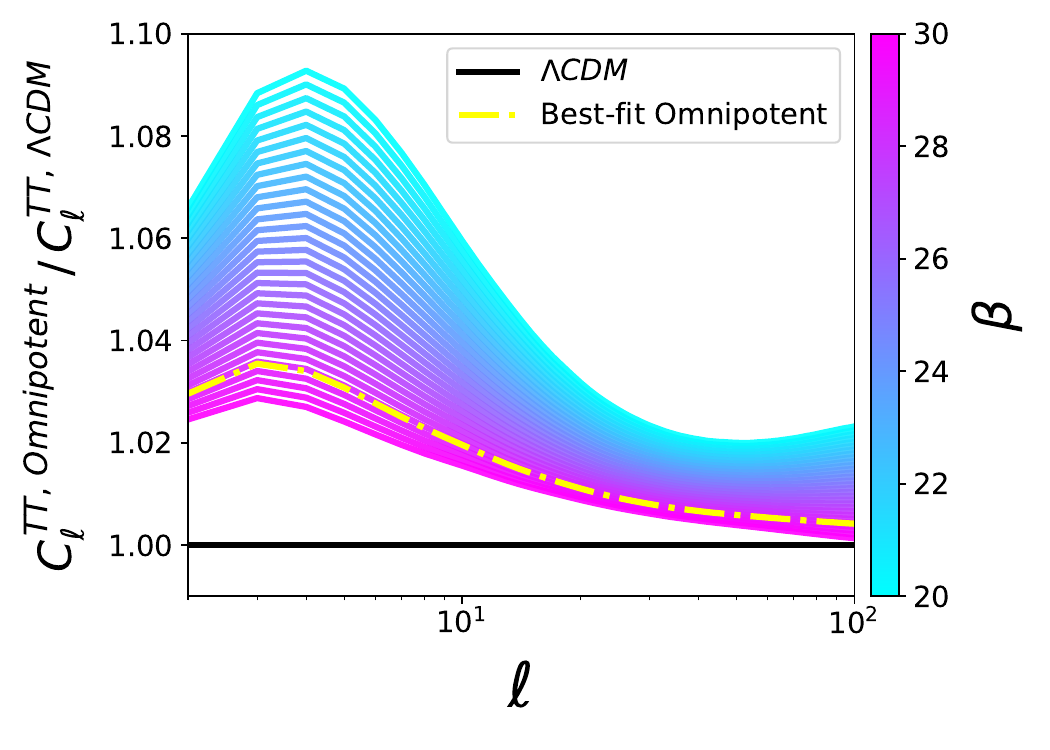}
    \caption{
Ratio of the CMB temperature power spectrum relative to the $\Lambda$CDM prediction, shown as $(C_\ell^{TT,\mathrm{Omnipotent}} / C_\ell^{TT,\Lambda\mathrm{CDM}})$, as a function of multipole moment $\ell$, for different values of the parameters $a_m$ (top), $\alpha$ (middle), and $\beta$ (bottom). 
The highlighted curves correspond to the best-fit parameter values.}
    \label{fig:C_TT_combined}
\end{figure}

For the Omnipotent DE model, we perform analogous one-parameter variations for $a_m$, $\alpha$, and $\beta$, keeping all other cosmological and model parameters fixed to their CMB+BAO best-fit values. As illustrated in Fig.~\ref{fig:C_TT_combined}, each parameter modifies the low-$\ell$ region in a distinct way:

\begin{itemize}
    \item \textbf{Top row:} Dependence of $C_l^{TT}$ on $a_m$ is nonmonotonic for a given $l$. This is most clearly seen at the lowest multipoles, where $C_l^{TT}$ decreases as $a_m$ moves away from 0, until around $a_m\sim0.6$, after which it begins to increase again. The overall trend shows an enhancement of the CMB power relative to $\Lambda$CDM for $0.7 \lesssim a_m \lesssim 1.0$. Despite the non-monotonic behavior that recovers the trend upwards at the smallest $a_m$ values, the power remains well below $\Lambda$CDM. This behavior reflects how the timing of the phantom crossing modulates the decay of gravitational potentials.

    \item \textbf{Middle row:} Varying $\alpha$ primarily rescales the amplitude of the deviation while keeping its shape nearly unchanged, confirming that $\alpha$ controls the overall strength of the omnipotent correction. The power is above the $\Lambda$CDM level for essentially all multipoles when $\alpha \gtrsim 12$, while for smaller values ($\alpha \lesssim 12$) a slight deficit appears only at high multipoles ($\ell \gtrsim 30$).

    \item \textbf{Bottom row:} Changes in $\beta$ modify the detailed multipole dependence of the spectrum. All curves for $20 \lesssim \beta \lesssim 30$ lie above $\Lambda$CDM across all multipoles, while smaller $\beta$ values produce a larger deviation above $\Lambda$CDM at all $\ell$, indicating that $\beta$ governs the asymmetry and curvature of the deviation pattern.
\end{itemize}

Altogether, these behaviors illustrate that the Omnipotent model has a rich phenomenology, in which each parameter leaves a characteristic imprint on the CMB, providing clear avenues for constraining the model with current and future data.

\subsection{Matter power spectrum}

\begin{figure*}[t]
\centering
    \includegraphics[width=0.48\linewidth]{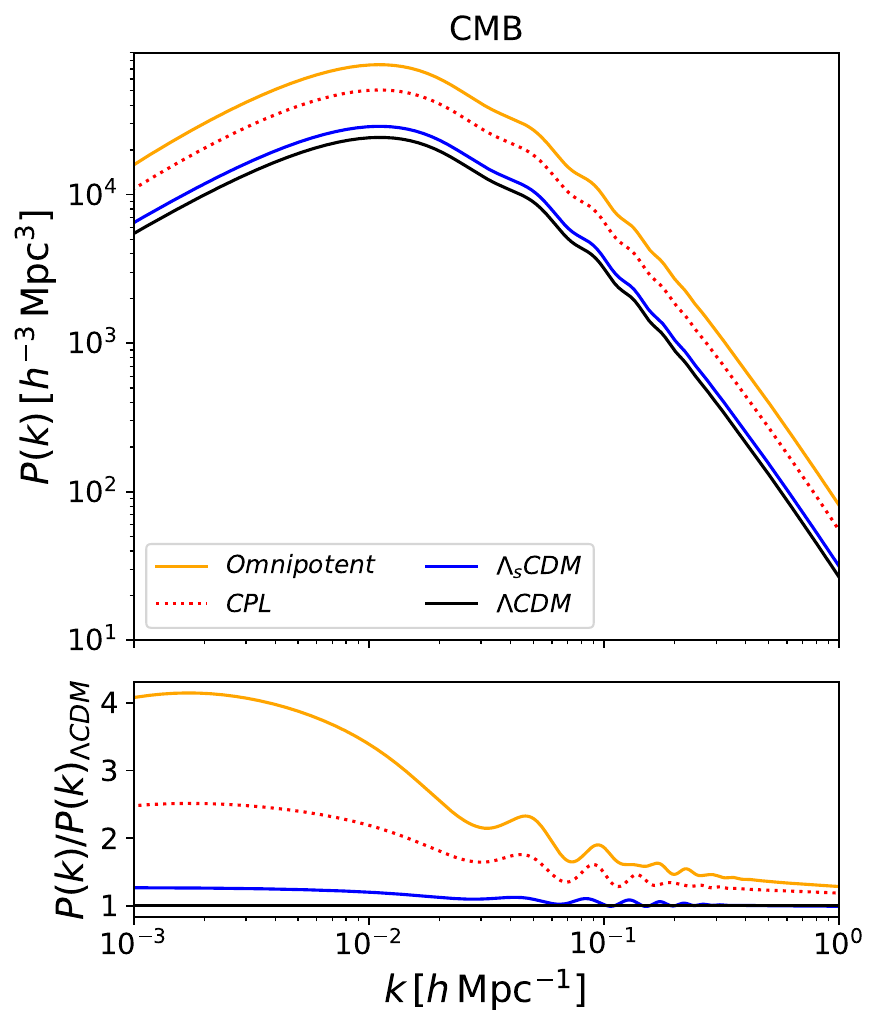}
    \includegraphics[width=0.48\linewidth]{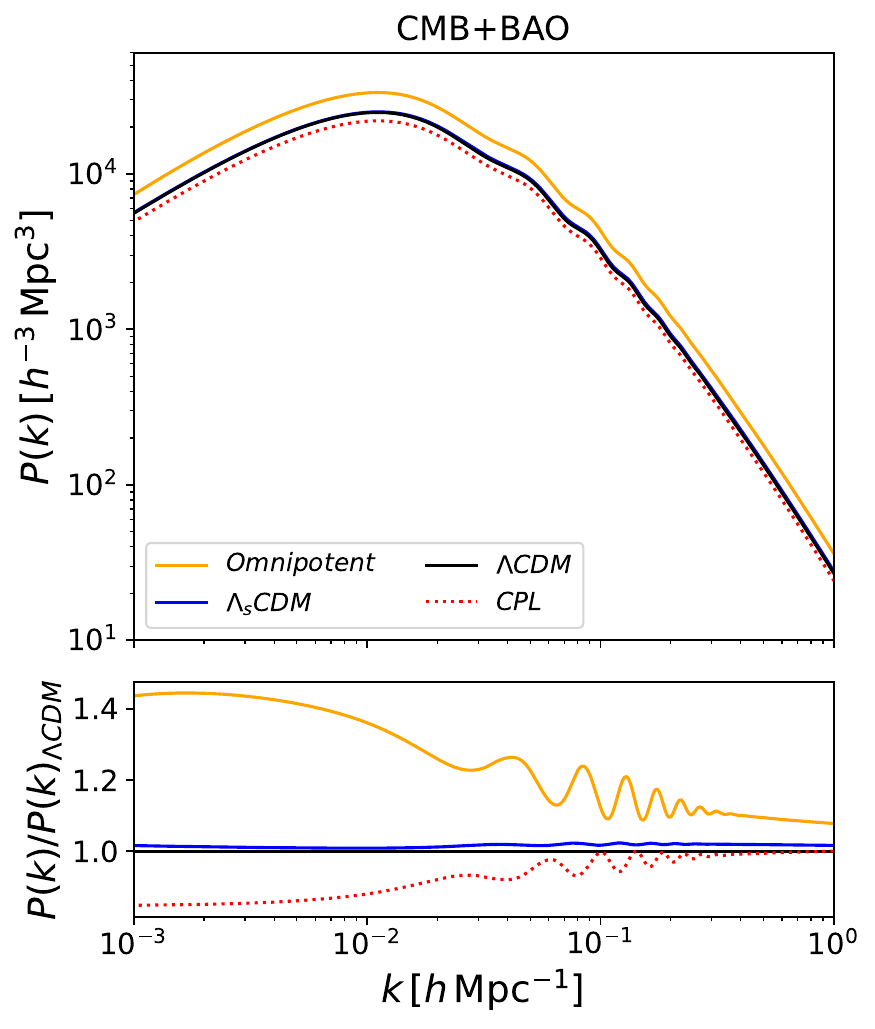}
    
    \caption{Matter power spectrum predictions for the dark energy models studied. The orange solid line corresponds to the Omnipotent DE model, the red dotted line to CPL, the blue solid line to $\Lambda_{\rm s}$CDM, and the black solid line to the standard $\Lambda$CDM model. The top panels display the matter power spectra, while the bottom panels show the corresponding ratios relative to $\Lambda$CDM to highlight the deviations more clearly. The left column uses best-fit parameters from the CMB-only analysis, whereas the right column shows results obtained from the combined CMB+BAO best-fit values.}
    \label{fig:mat}
\end{figure*}

We complement the CMB analysis with the matter power spectrum, $P(k)$, evaluated at redshift $z = 0$. 
Fig.~\ref{fig:mat} shows the predictions for all four models, using the best-fit cosmologies from both the CMB-only and CMB+BAO analyses. 
The spectra cover the range $10^{-3} < k < 1 \, [h\,\mathrm{Mpc}^{-1}]$ and are computed with \texttt{CAMB}, including non-linear corrections through the \texttt{Halofit} prescription.\footnote{For dedicated $N$-body simulations of the $\Lambda_{\rm s}$CDM model and the corresponding nonlinear matter power spectrum, see Ref.~\cite{Akarsu:2025nns}.}

The left top panel shows that, despite being nearly indistinguishable in the CMB temperature anisotropy spectra (see Subsection~\ref{sec:TT}), the dark energy models produce noticeably different matter power spectra when using CMB-only best fits. 
This discrepancy arises from parameter degeneracies in the CMB data, which limit the precision with which the overall clustering amplitude can be determined. 
Among the models, Omnipotent DE (orange solid line) exhibits the highest power across all scales, indicating enhanced structure growth, followed by the CPL (red dashed line) with a moderately smaller amplitude. 
The $\Lambda_{\mathrm{s}}$CDM prediction (blue solid line) lies close to the $\Lambda$CDM reference (black solid line), which yields the lowest overall power.

The top-right panel highlights the strong constraining power of BAO data. When the BAO measurements are combined with the CMB likelihood, degeneracies in the growth-related parameters are efficiently broken. As a result, the $\Lambda_{\mathrm{s}}$CDM spectrum becomes nearly indistinguishable from $\Lambda$CDM across the entire $k$-range, with the two curves overlapping throughout the plot. The CPL spectrum remains below $\Lambda$CDM over most of the range, becoming visually close to it only at high wavenumbers ($k \gtrsim 0.3$), while the Omnipotent model stays above $\Lambda$CDM at all scales, exhibiting a particularly large enhancement for $k \lesssim 0.01$. For clarity, the bottom panels of both plots show the ratio $(P(k)_{\mathrm{Model}} \,/\, P(k)_{\Lambda \mathrm{CDM}})$, which more clearly highlights deviations from $\Lambda$CDM across all scales.

This comparison demonstrates that models producing nearly identical CMB anisotropies can still predict distinct growth histories on different scales. 
Incorporating BAO data, which tightly constrain the late-time expansion rate, effectively anchors the amplitude of the matter power spectrum and restricts the freedom of dark energy models to deviate from $\Lambda$CDM behaviour.

\section{Integrated Sachs-Wolfe effect}\label{sec:ISW}

The ISW effect arises from the coupling between the expansion history of the Universe and the time evolution of gravitational potentials associated with large-scale structures. It is a secondary CMB anisotropy generated along the line of sight and provides a direct probe of the late-time dynamics of cosmic acceleration.

After recombination, CMB photons propagate freely toward the observer. During the matter-dominated era, gravitational potentials remain approximately constant on linear scales, so no late-time ISW contribution is produced. When the cosmic expansion departs from matter domination, whether due to a positive cosmological constant, a dynamical dark energy component, or an exotic sector with negative energy density,\footnote{Note that in both models considered here that allow negative DE densities, previous studies have shown that, when their parameters are constrained by the data, the regime with negative DE density ends at $z \gtrsim 1.5$ during matter domination.} the gravitational potentials begin to evolve in time, leading to additional temperature anisotropies through the ISW effect~\cite{Sachs:1967er,Hu:1995hf}.

Although the ISW signal is subdominant compared with the primary CMB anisotropies, it leaves a distinct imprint on large angular scales (low multipoles, $\ell \lesssim 100$)~\cite{Schaefer:2005up}. In this work, we use multiple observational probes of the ISW effect (its imprint on the CMB temperature power spectrum, the ISW–galaxy cross-correlation, and the lensing–ISW bispectrum) to test how exotic dark energy models with negative or sign-changing densities modify the late-time evolution of gravitational potentials.

\subsection{Theory}

As CMB photons traverse evolving gravitational potentials, they experience successive blueshifts when falling into potential wells and redshifts when climbing out. 
If the depth of these potentials changes over time, the energy gained and lost by photons along the line of sight does not cancel exactly, leading to a net temperature anisotropy in the observed CMB. 
This contribution, known as the Integrated Sachs–Wolfe effect, can be expressed as~\cite{Schaefer:2008qs}:
\begin{equation}
	\Theta_{\mathrm{ISW}} = \left( \dfrac{\Delta T}{T_{\mathrm{CMB}}} \right)_{\mathrm{ISW}} = 
	- \dfrac{2}{c^3} \int_0^{\chi_H} a^2 H(a) \, \dfrac{\partial \phi}{\partial a} \, d\chi,
\end{equation}
where $T_{\mathrm{CMB}} = 2.725\,\mathrm{K}$ is the mean CMB temperature, $\phi$ is the gravitational potential, $c$ is the speed of light, and $\chi$ is the comoving distance to the observer. 
The comoving distance as a function of the scale factor $a$ is given by:
\begin{equation}
	\chi(a) = \int_a^1 \dfrac{c \, da'}{a'^2 H(a')}.
\end{equation}

In this picture, CMB photons are affected by two competing mechanisms: the overall redshift due to cosmic expansion and the gravitational redshift and blueshift caused by evolving matter inhomogeneities. 
These competing effects lead to small but measurable variations in the observed CMB temperature~\cite{Wang:2016lxa,Boughn:2003yz}.

The corresponding angular power spectra of the ISW temperature fluctuations and their cross-correlation with large-scale structure tracers can be written, in the Limber approximation, as~\cite{Schaefer:2008qs,Muir:2016veb,Afshordi:2003xu,Yengejeh:2022tpa}:
\begin{equation}
\begin{split}
C^{TT}_{\mathrm{ISW}} (\ell) &= \int_0^{\chi_H} \dfrac{W_T^2(\chi)}{\chi^2} \, \dfrac{H_0^4}{k^4} \,
P\!\left(k = \dfrac{\ell + 1/2}{\chi}\right) d\chi, \\
C^{Tg}_{\mathrm{ISW}} (\ell) &= \int_0^{\chi_H} \dfrac{W_T(\chi) W_g(\chi)}{\chi^2} \, \dfrac{H_0^2}{k^2} \,
P\!\left(k = \dfrac{\ell + 1/2}{\chi}\right) d\chi,
\end{split}
\end{equation}
where $W_T$ and $W_g$ denote the ISW and galaxy window functions, respectively. 
The galaxy window function depends on the survey selection and redshift distribution through:
\begin{equation}
W_g(z) \propto f_{\mathrm{survey}}(z) \equiv b(z) \, \dfrac{dn}{dz},
\label{W_g}
\end{equation}
where $b(z)$ is the (scale-independent) galaxy bias and $dn/dz$ is the normalized redshift distribution of the sources.

\begin{figure*}[t]
    \centering
    \includegraphics[width=0.48\textwidth]{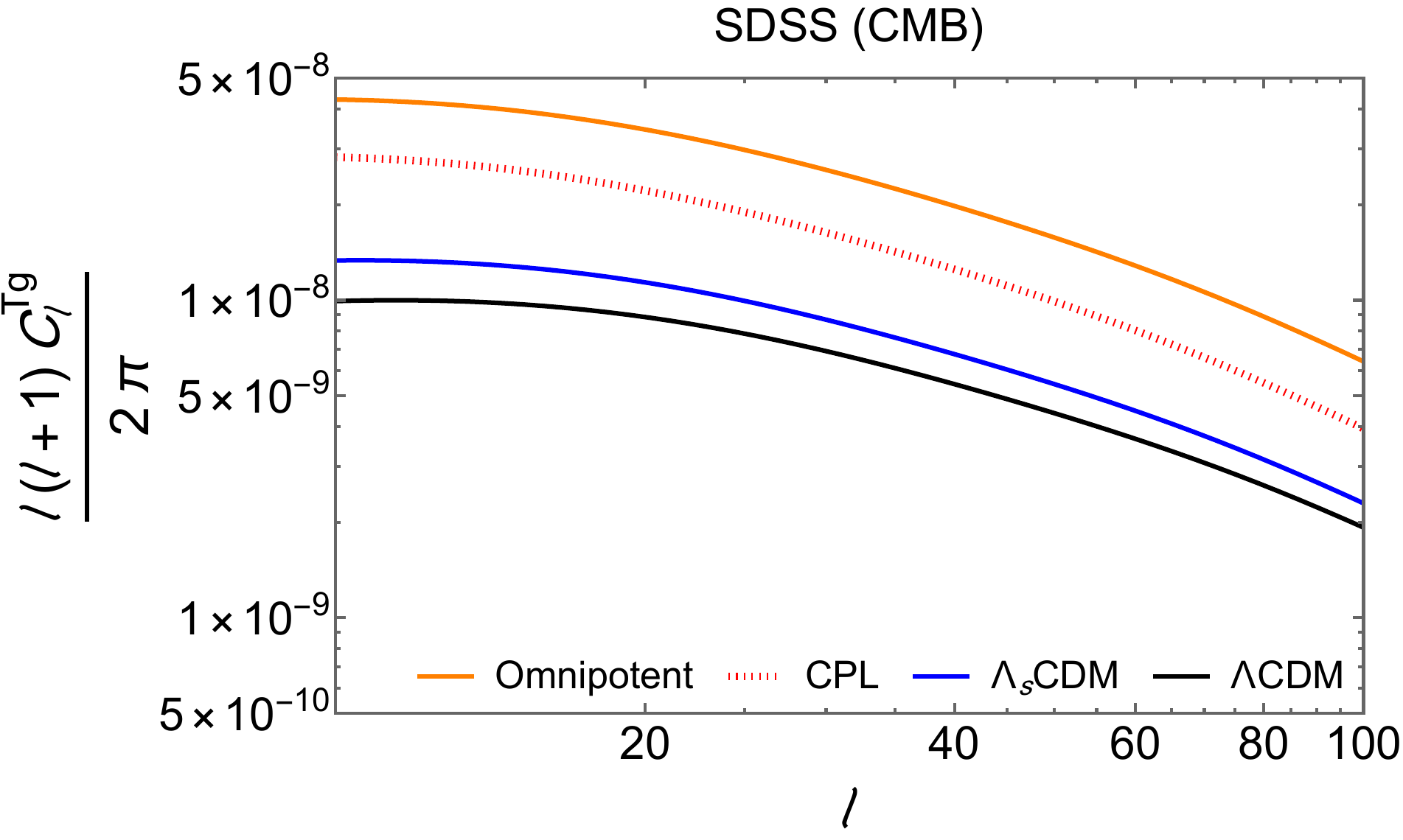}
    \includegraphics[width=0.48\textwidth]{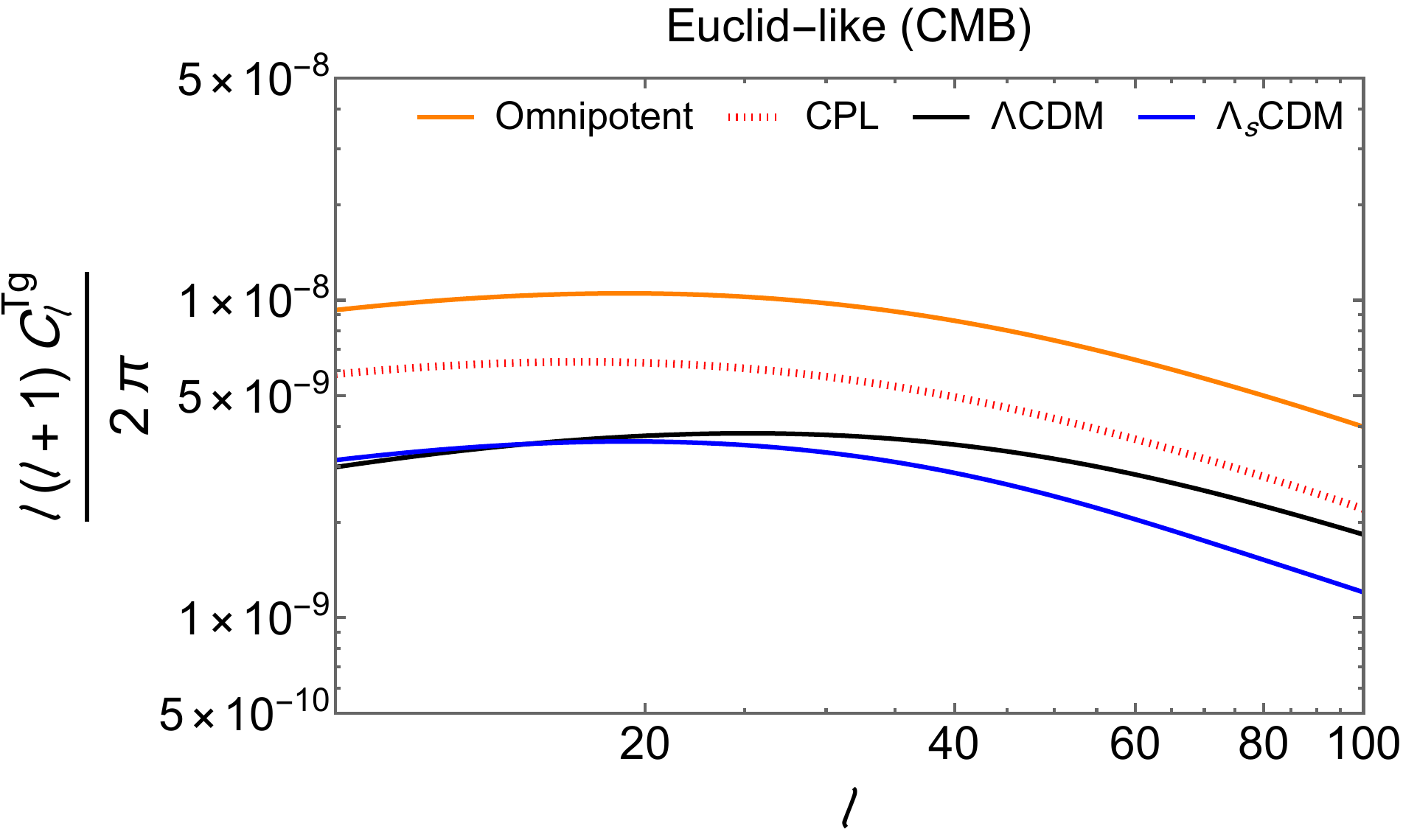}
\vspace{0.5em}
 \includegraphics[width=0.48\textwidth]{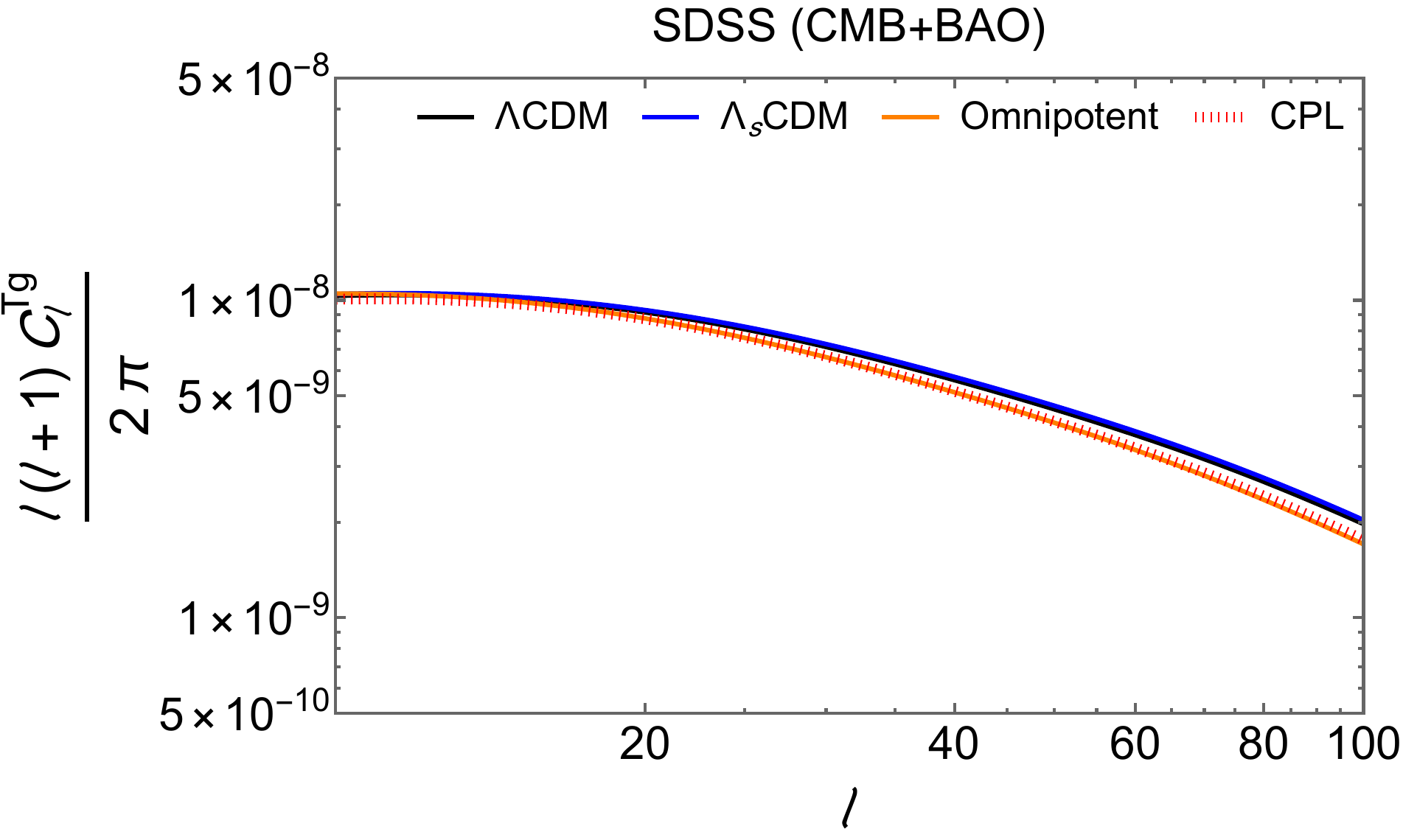} \includegraphics[width=0.48\textwidth]{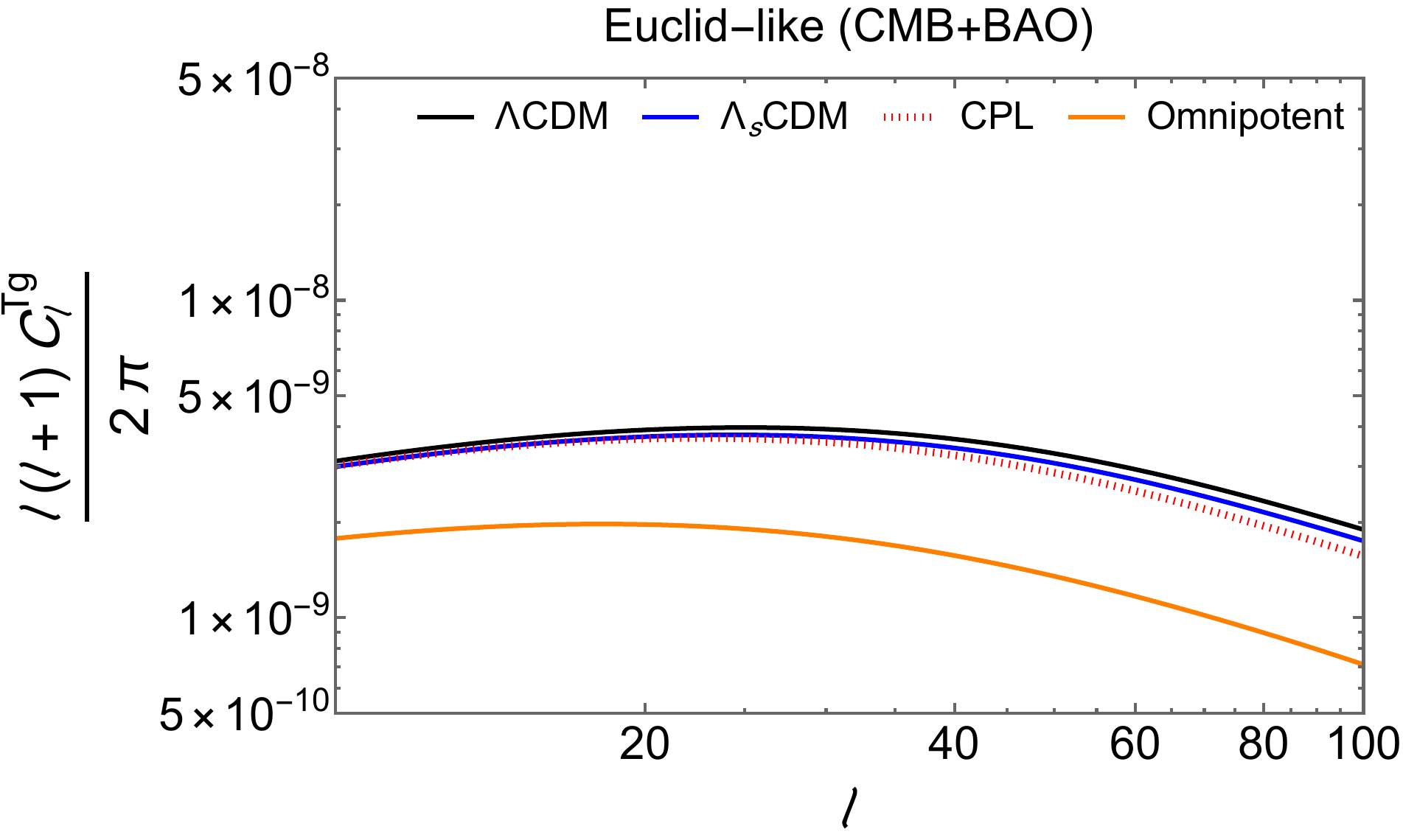}
    \vspace{0.5em} 
    \includegraphics[width=0.48\textwidth]{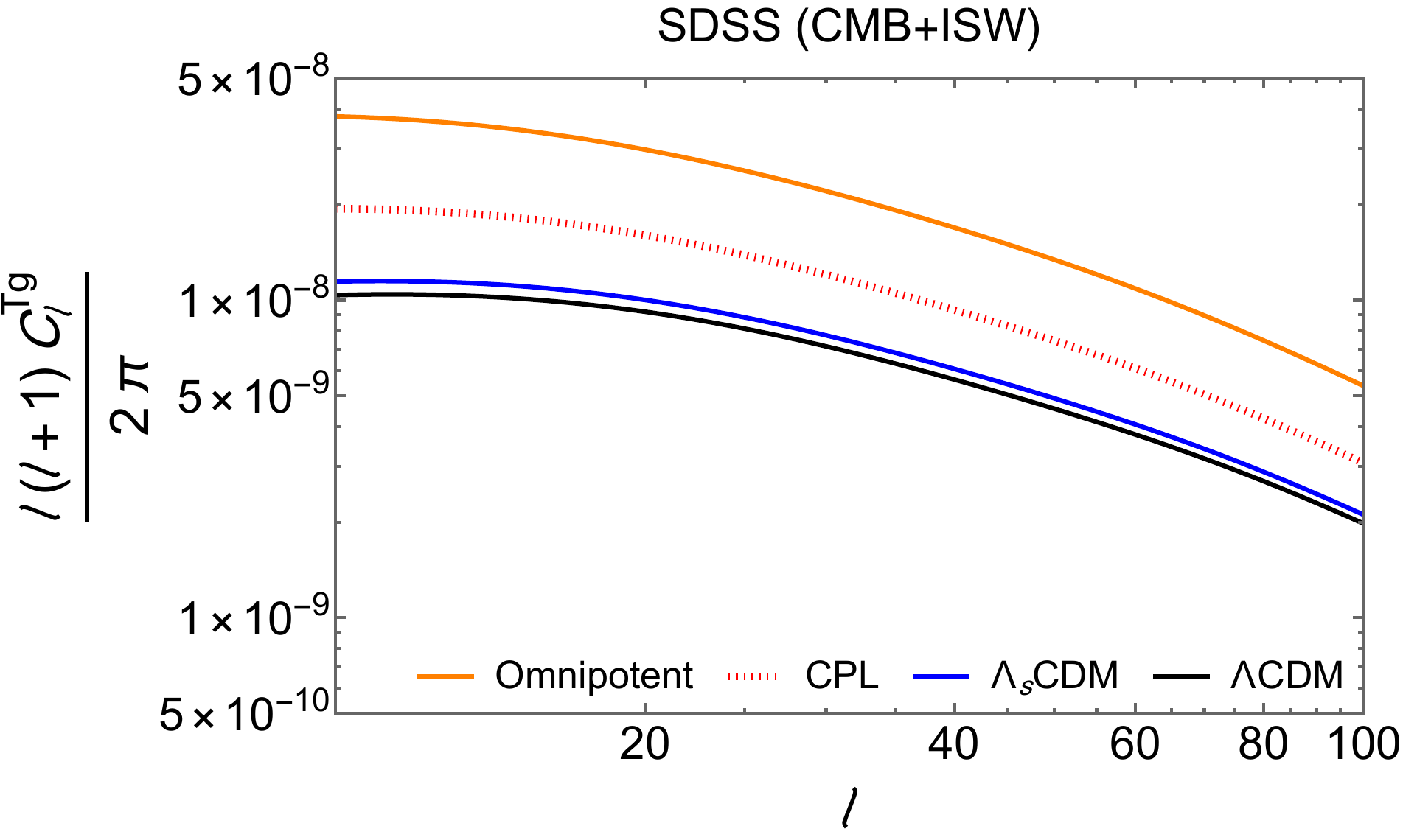}
    \includegraphics[width=0.48\textwidth]{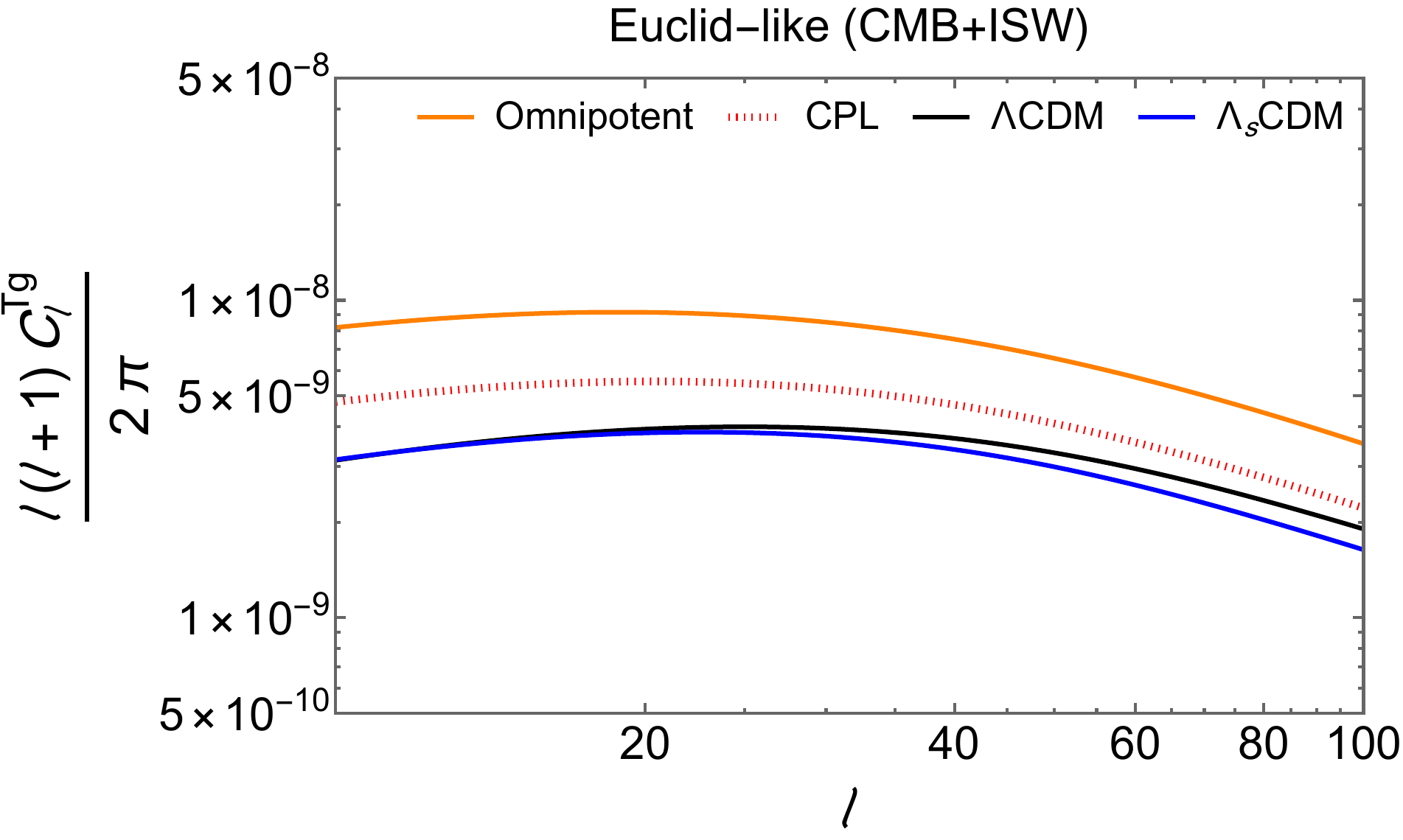}
    \caption{
The ISW–galaxy cross-power spectrum, $C_\ell^{Tg}$, for SDSS (left column) and a Euclid-like survey (right column), computed using the best-fit cosmological parameters from the CMB-only (top row), CMB+BAO (middle row), and CMB+ISW (bottom row) analyses. 
Predictions are shown for $\Lambda$CDM (black solid line), $\Lambda_{\mathrm{s}}$CDM (blue solid line), Omnipotent DE (orange solid line), and CPL DE (red dotted line) models. 
}
    \label{fig:C_Tg_combined}
\end{figure*}

\subsection{Surveys}
\label{sec:surveys}

While the ISW contribution to the CMB temperature anisotropies is subtle and difficult to isolate on its own, its cross-correlation with galaxy surveys provides a powerful probe of the late-time evolution of gravitational potentials. 
The effectiveness of this approach depends critically on the redshift distribution of the tracers used to map the large-scale structure. 
This distribution, encoded in the galaxy selection function of Eq.~\ref{W_g}, determines the redshift range over which a survey is most sensitive to the ISW effect and therefore its capacity to distinguish between different cosmological models.

In this work, we adopt redshift distributions representative of two surveys: the Sloan Digital Sky Survey (SDSS) and an Euclid-like large-scale structure survey, the latter based on expected survey specifications. 
The corresponding redshift-dependent selection functions are defined as~\cite{Ho:2008bz,Velten:2015qua,Margon:1998vu,Peiris:2000kb,Planck:2015fcm,Martinet:2015wza,Weaverdyck:2017ovf}:
\begin{equation} \label{eq:RDF-SDSS}
f_{\mathrm{SDSS}}(z) = b_{\mathrm{eff}} \, 
\dfrac{a_*}{\Gamma \!\left( \frac{m+1}{a_*} \right)} 
\dfrac{z^m}{z_*^{m+1}} 
\exp \!\left[-\left(\dfrac{z}{z_*}\right)^{a_*}\right],
\end{equation}
\begin{equation} \label{eq:RDF-Euclid}
f_{\mathrm{Euclid\text{-}like}}(z) = b_{\mathrm{eff}} \, 
\dfrac{3}{2 z_*^3} \, z^2 \exp \!\left[-\left(\dfrac{z}{z_*}\right)^{3/2}\right],
\end{equation}
where $\Gamma(x)$ denotes the Gamma function, and $b_{\mathrm{eff}}$, $z_*$, $a_*$, and $m$ are free parameters. 
The best-fit values adopted in our analysis are listed in Table~\ref{tab:Redshift_Best}.

\begin{table}[htbp]
    \centering
    \begin{tabular}{c|c|c|c|c}
    \hline\hline
    Survey & $b_{\mathrm{eff}}$ & $z_*$ & $a_*$ & $m$ \\
    \hline\hline
    SDSS & $1.00$ & $0.113$ & $1.197$ & $3.457$ \\
    \hline
    Euclid-like & $1.00$ & $0.700$ & $-$ & $-$ \\
    \hline \hline
    \end{tabular}
    \caption{
Best-fit values of the redshift distribution parameters adopted for the SDSS and Euclid-like galaxy surveys.
}
    \label{tab:Redshift_Best}
\end{table}

Fig.~\ref{fig:C_Tg_combined} presents the ISW–galaxy cross-power spectrum, $C_\ell^{Tg}$, as a function of multipole moment, $\ell$, in the range $10 \leq \ell \leq 100$, computed using the best-fit cosmological parameters from Tables~\ref{tab:CMB}–\ref{tab:CMB+ISW} and the redshift distribution parameters listed in Table~\ref{tab:Redshift_Best}. 
The six panels correspond to two galaxy surveys—SDSS (left column) and a Euclid-like survey (right column)—and three sets of cosmological constraints: CMB-only (top row), CMB+BAO (middle row), and CMB+ISW (bottom row). 
Each panel shows predictions for the $\Lambda$CDM (black solid line), $\Lambda_{\mathrm{s}}$CDM (blue solid line), Omnipotent DE (orange solid line), and CPL (red dotted line) models, using the respective redshift distributions and bias parameters of each survey. 
This multi-panel comparison highlights how different dark energy scenarios modify the correlation between the CMB temperature and the large-scale structure at late times. 
It enables us to disentangle the impact of the late-time expansion history from that of the evolving gravitational potentials, thereby clarifying the cosmological mechanisms driving the ISW signal.

\begin{itemize}

    \item \textbf{Survey-dependent sensitivity to late-time evolution}:  
    The differences between the SDSS and Euclid-like results arise from their distinct redshift distributions.  
    SDSS predominantly probes the very low-redshift Universe, where the ISW signal is strongest, and therefore exhibits larger model-to-model variations whenever dark energy significantly affects the decay of gravitational potentials near the present epoch.  
    In contrast, the Euclid-like survey extends to higher redshifts, where the ISW effect becomes weaker, resulting in smaller differences among models.  
    The relative similarity between models in the Euclid-like panels thus reflects a milder late-time modification integrated over a broader redshift range.

    \item \textbf{Breaking late-time degeneracies with BAO constraints}:  
    The CMB+BAO results highlight the crucial role of BAO data in constraining the expansion history at intermediate and low redshifts.  
    Once BAO information is included, models that primarily alter background distances or the overall growth amplitude are driven toward $\Lambda$CDM-like behaviour.  
    This explains why both the CPL and $\Lambda_{\mathrm{s}}$CDM predictions become nearly indistinguishable from $\Lambda$CDM for both surveys.  
    The Omnipotent DE model, however, still shows visible deviations, indicating that its free parameters modify the growth rate and the scale dependence of the ISW kernel more strongly than the other models.

    \item \textbf{Role of ISW data in shaping best-fit parameters}:  
    In the CMB+ISW case, the additional likelihood corresponds to the \textit{Planck} lensing–temperature cross-correlation measurement~\cite{Carron:2022eum}, which provides a direct observational detection of the late-time Integrated Sachs–Wolfe signal. 
    Including this dataset refines the sensitivity to the decay rate of gravitational potentials at low redshifts.  
    For SDSS, which traces structures near the ISW kernel peak, the $\Lambda_{\mathrm{s}}$CDM and $\Lambda$CDM spectra remain closely aligned, implying that the best-fit transition redshift $z^{\dagger}$ corresponds to a nearly $\Lambda$CDM-like potential evolution.  
    In contrast, the Omnipotent DE model predicts an enhanced cross-correlation amplitude across multipoles, pointing to a stronger present-day potential decay.  
    For the Euclid-like case, deviations are more scale-dependent.

\end{itemize}

\begin{figure}[htbp]
    \centering
    \includegraphics[width=0.98\linewidth]{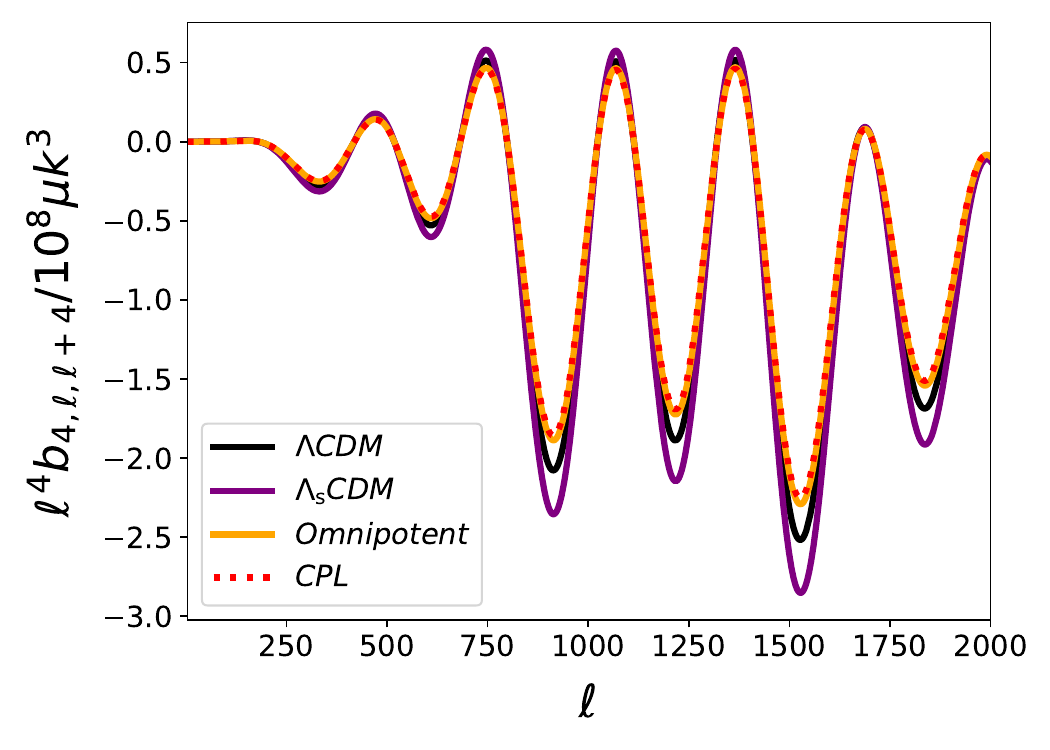}
    \caption{
Reduced bispectrum $b_{\ell_1\ell_2\ell_3}$ for the different dark energy models, computed using the CMB-only best-fit cosmological parameters. 
The black solid line represents the $\Lambda$CDM model, the red dotted line corresponds to the CPL model, the blue solid line to the $\Lambda_{\mathrm{s}}$CDM model, and the orange solid line to the Omnipotent DE model.
}
    \label{fig:bispectrum_all}
\end{figure}
\begin{figure}[htbp]
    \centering
    \includegraphics[width=0.95\linewidth]{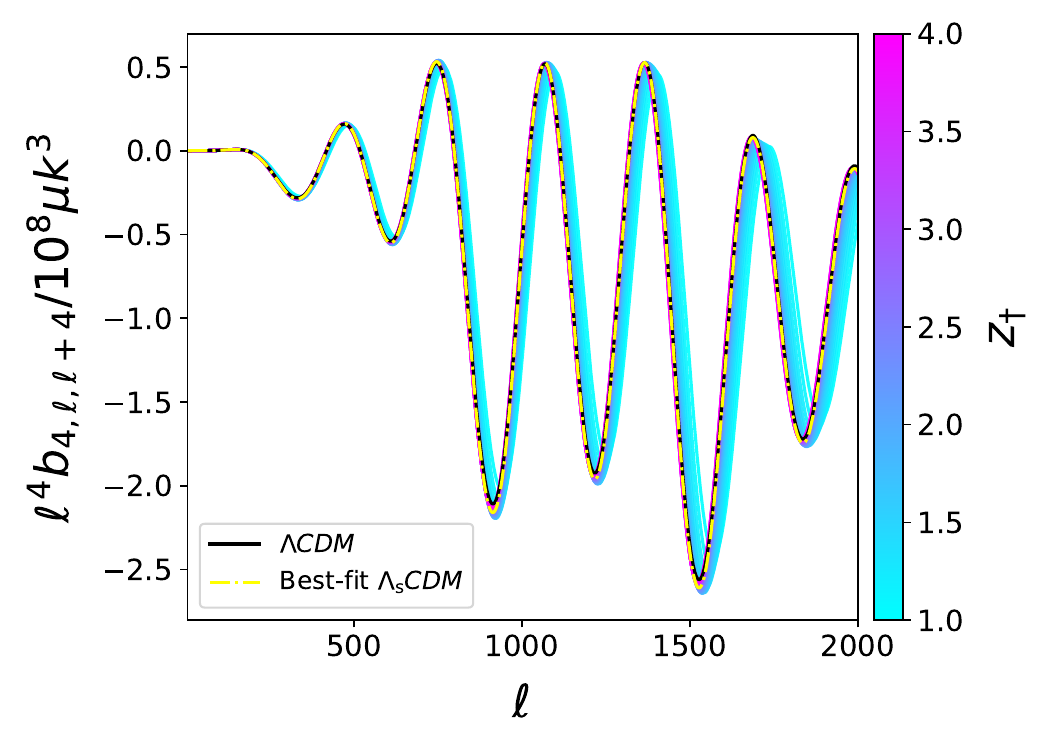}
    \caption{
Reduced bispectrum $b_{\ell_1\ell_2\ell_3}$ for the $\Lambda_{\mathrm{s}}$CDM dark energy model, shown as a function of multipole moment $\ell$ for different values of the transition redshift parameter $z^{\dagger}$. 
The highlighted line corresponds to the best-fit result.
}
    \label{fig:bispectrum_ZDAG}
\end{figure}
\begin{figure}[htbp]
    \centering
    \includegraphics[width=0.95\linewidth]{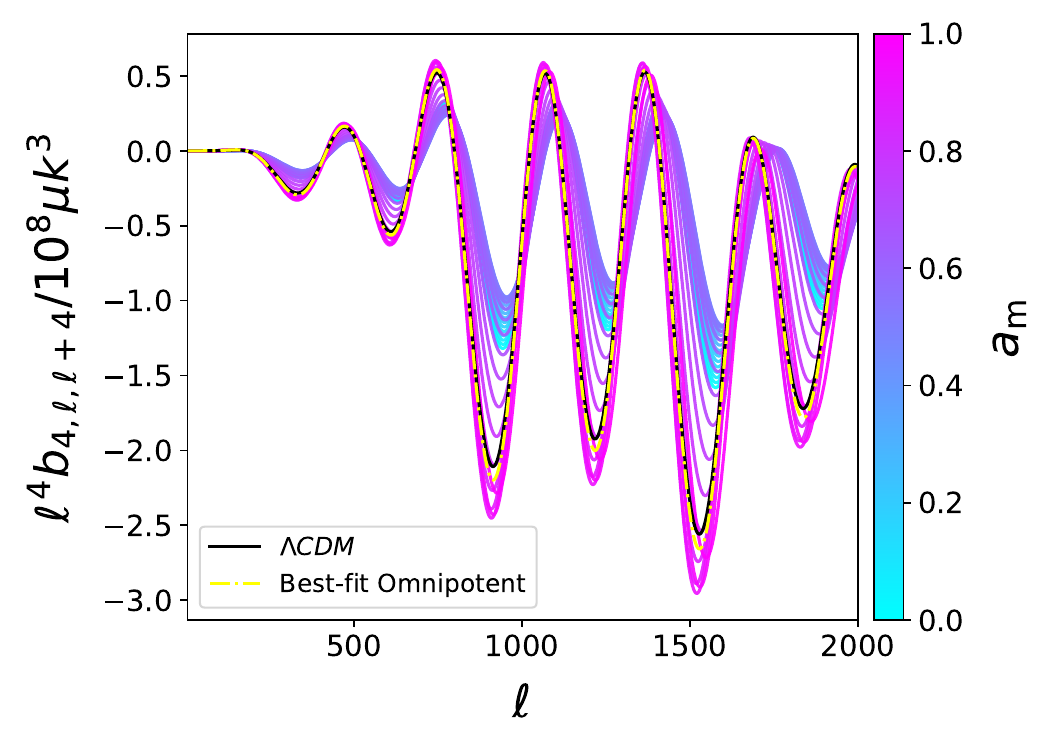}
        \includegraphics[width=0.95\linewidth]{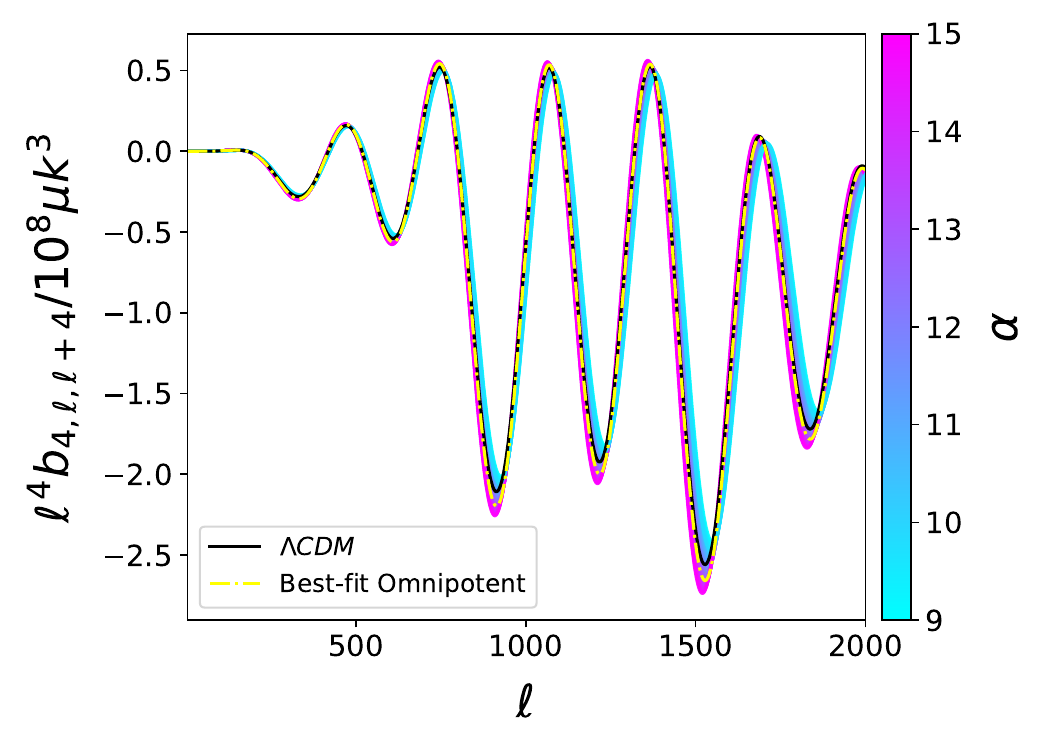}
          \includegraphics[width=0.95\linewidth]{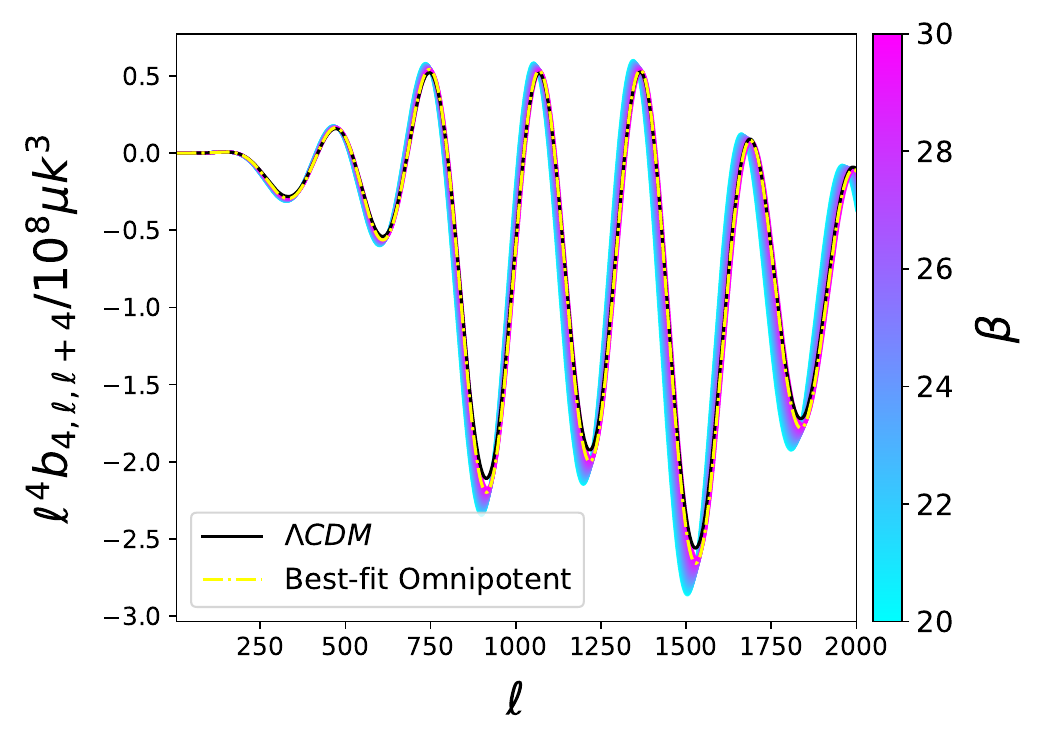}
    \caption{
Reduced bispectrum $b_{\ell_1\ell_2\ell_3}$ for the Omnipotent DE model, shown as a function of multipole moment $\ell$ for variations in the parameters $a_m$ (top), $\alpha$ (middle), and $\beta$ (bottom). 
The highlighted line corresponds to the best-fit result.}
    \label{fig:bispectrum_CMB}
\end{figure}

\section{Measuring The Lensing-ISW Bispectrum}
\label{sec:bispectrum}

The ISW effect and weak gravitational lensing both originate from the large-scale gravitational potential at late times: the ISW effect traces its temporal decay, while lensing maps its spatial distribution along the line of sight~\cite{Cooray:1999kg, Smith:2006ud}. Since both depend on the evolution of the potential driven by dark energy, their correlation encodes valuable information about the late-time expansion history. This correlation generates a characteristic nonzero CMB bispectrum signal, known as the lensing–ISW bispectrum. Measuring this signal provides a higher-order statistical probe of the late-time Universe, offering a complementary test of the dark energy models considered in this work. By comparing the predicted lensing–ISW bispectra across models, we aim to identify distinctive signatures that could help discriminate between them using CMB observations.

Unlike the primary bispectrum produced by primordial non-Gaussianity, the lensing–ISW bispectrum arises from secondary anisotropies. Its shape and amplitude are highly model dependent and sensitive to late-time physics, which in principle makes it a powerful observable for breaking degeneracies between cosmological models that otherwise yield similar fits to two-point functions.

CMB photons are affected by the integrated gravitational potential along their path~\cite{Boughn:2003yz}, while weak lensing further distorts their trajectories. Examining the third-order statistical correlation~\cite{Philcox:2023uwe} between the lensing potential and the ISW temperature fluctuations—both sourced by the Weyl potential—allows one to quantify the lensing–ISW bispectrum. In this context, the observed CMB anisotropy in a given direction is not measured exactly along the unperturbed line of sight $\hat{n}$, but along the deflected direction $\hat{n} + \nabla\phi(\hat{n})$, where $\phi$ is the lensing potential. As a result, both the temperature anisotropy and the lensing potential are modified~\cite{DiValentino:2012yg, Fergusson:2010dm}:
\begin{equation}\label{eq:CMB-ANISOTROPY}
    \delta\tilde{T}(\hat{n}) = \delta T(\hat{n} + \nabla\phi(\hat{n})),
\end{equation}
\begin{equation}
    \phi(\hat{n}) = \sum_{\ell m} \phi_{\ell m} Y_{\ell m}(\hat{\mathbf{n}}).
\end{equation}

Here, $\delta \tilde T(\hat{n})$ denotes the lensed CMB anisotropy, while $\delta T(\hat{n})$ is the unlensed ISW contribution. From Eq.~\ref{eq:CMB-ANISOTROPY}, long-wavelength ISW modes correlate with short-wavelength lensing modes, inducing a nonzero reduced bispectrum $b_{\ell_1 \ell_2 \ell_3}$ in harmonic space~\cite{Spergel:1999xn, Goldberg:1999xm}:
\begin{eqnarray}
  b_{\ell_1\ell_2\ell_3} = h_{\ell_1\ell_2\ell_3}^{-1} B_{\ell_1\ell_2\ell_3} 
  = M \times C^{T\phi}_{\ell_2} C^{TT}_{\ell_3} + 5~\mathrm{perm.},
\end{eqnarray}
where
\begin{equation}
M = \frac{-\ell_1(\ell_1+1) + \ell_2(\ell_2+1) + \ell_3(\ell_3+1)}{2},
\end{equation}
and $C^{TT}_{\ell}$ is the temperature power spectrum, while $C^{T\phi}_{\ell} = \langle \phi^*_{\ell m} a_{\ell m} \rangle$ denotes the cross-correlation between the lensing potential and the temperature anisotropy~\cite{Hu:2012td}. 
Numerically, it is more convenient to work with the reduced bispectrum $b_{\ell_1\ell_2\ell_3}$, which removes the geometrical factor $h_{\ell_1\ell_2\ell_3}$ from the full bispectrum expression~\cite{Fergusson:2010dm}.

Fig.~\ref{fig:bispectrum_all} shows the reduced lensing-ISW bispectrum, $b_{\ell_1\ell_2\ell_3}$, as a function of multipole moment ($4 \leq \ell \leq 2000$) for the four dark energy models, computed using the CMB-only best-fit parameters. The bispectra display oscillatory features, with alternating maxima and minima in amplitude. These oscillations arise from the coupling between long-wavelength modes associated with the late-time ISW effect and short-wavelength modes generated by gravitational lensing. Models that alter the expansion and growth histories at late or intermediate redshifts modify this coupling strength, thereby enhancing or suppressing the bispectrum amplitude.
Although the models are nearly indistinguishable at the level of the CMB temperature power spectrum, they yield distinct predictions for the lensing--ISW bispectrum. 
The $\Lambda_{\rm s}$CDM model (blue solid line) exhibits the largest oscillation amplitude, with both higher peaks and deeper troughs than $\Lambda$CDM, reflecting a stronger correlation between ISW and lensing signals. 
The Omnipotent DE model (orange solid line) follows a similar pattern but with a slightly reduced amplitude, while the CPL model (red dotted line) shows smaller variations, consistent with a weaker ISW-lensing coupling. 
These differences highlight the potential of the bispectrum as a sensitive probe of late-time gravitational dynamics and a discriminator among dark energy scenarios that otherwise produce nearly identical two-point statistics.

In Fig.~\ref{fig:bispectrum_ZDAG}, we show the reduced bispectrum of the $\Lambda_{\rm s}$CDM model for different values of the transition redshift $z^{\dagger}$, keeping all other cosmological parameters fixed to the CMB+BAO best-fit values. As $z^{\dagger}$ decreases, the transition in the cosmological constant occurs at later times and the lensing-ISW bispectrum is more strongly distorted: the positions of the extrema shift towards higher multipoles, the negative wells become more pronounced, and the positive peaks are slightly suppressed with respect to the $\Lambda$CDM case. For larger values of $z^{\dagger}$, the bispectrum progressively approaches the $\Lambda$CDM behaviour, as expected in the limit $z^{\dagger} \to \infty$ where the sign switch effectively moves out of the observable range. This trend is in qualitative agreement with the recent analysis of the lensing-ISW bispectrum in $\Lambda_{\rm s}$CDM cosmologies presented in Ref.~\cite{Forconi:2025gwo}.

In Fig.~\ref{fig:bispectrum_CMB}, we present the reduced bispectrum of the Omnipotent DE model as a function of multipole moment ($3 < \ell < 2100$), illustrating the effect of varying the parameters $a_m$, $\alpha$, and $\beta$, while keeping all other cosmological and model parameters fixed to their CMB+BAO best-fit values. The three panels correspond to variations in $a_m$ (top), $\alpha$ (middle), and $\beta$ (bottom).
The bispectrum exhibits oscillatory features whose amplitude and position are strongly influenced by $a_m$. As shown in the top panel, decreasing $a_m$ shifts the peaks and troughs of the bispectrum systematically toward higher multipoles, indicating that the ISW–lensing correlation is pushed to smaller angular scales. At the same time, the amplitudes of both peaks and wells are reduced, with a more pronounced suppression in the negative minima. Among the three parameters, $a_m$ produces the most noticeable changes in the overall scale at which the signal is modulated.
The middle panel shows the effect of varying $\alpha$. In this case, the sensitivity is milder: larger values of $\alpha$ increase the amplitude of both peaks and wells, while smaller values reduce their amplitude and shift the extrema slightly toward higher multipoles. Thus, $\alpha$ primarily controls the overall strength of the bispectrum signal, with only modest impact on the precise multipole positions of the features.
The bottom panel displays the impact of changing $\beta$. Here the behaviour is qualitatively opposite to that of $\alpha$: increasing $\beta$ tends to suppress the amplitude of the peaks and wells, whereas smaller values enhance them and shift the oscillatory pattern in the opposite direction in multipole space compared to $\alpha$ variations. Overall, while the bispectrum shape remains broadly similar, the parameters $a_m$, $\alpha$, and $\beta$ modulate its amplitude and effective scale in complementary ways, providing additional lever arms to distinguish Omnipotent DE from $\Lambda$CDM using higher-order CMB statistics.

To summarize, the reduced lensing–ISW bispectrum provides a higher-order statistical probe of the late-time Universe, capturing the non-Gaussian signatures generated by the coupling between evolving large-scale gravitational potentials and weak lensing. Although this signal is secondary in origin and distinct from primordial non-Gaussianity, it remains highly sensitive to the dynamics of dark energy. Its shape and amplitude are intrinsically model-dependent and can offer discriminatory power beyond that of two-point correlations, making it a valuable complement to the $C_\ell^{TT}$ and $C_\ell^{Tg}$ analyses discussed earlier.

\begin{table*}
\begin{center}
\resizebox{0.85\textwidth}{!}{%
\begin{tabular}{c||c|c|c|c}
\hline\hline
\textbf{Parameters} & \textbf{CMB} & \textbf{CMB+ISW}&\textbf{CMB+BAO} & \textbf{CMB+BAO+ISW} \\
\hline\hline
\textbf{\boldmath$\log(10^{10} A_\mathrm{s})$} & $ 3.036\pm 0.014 $ & $ 3.037\pm 0.015 $ & $ 3.040\pm 0.014 $ & $ 3.041\pm 0.014 $ \\ 
\textbf{\boldmath$n_\mathrm{s}$} & $ 0.9672\pm 0.0042 $ & $ 0.9674\pm 0.0041 $ & $ 0.9654\pm 0.0039 $ & $ 0.9653\pm 0.0038 $ \\ 
\textbf{\boldmath$100\theta_\mathrm{MC}$} & $ 1.04102\pm 0.00031 $ & $ 1.04101\pm 0.00031 $ & $ 1.04091\pm 0.00030 $ & $ 1.04092\pm 0.00030 $ \\ 
\textbf{\boldmath$\Omega_\mathrm{b} h^2$} & $ 0.02244\pm 0.00015 $ & $ 0.02244\pm 0.00015 $ & $ 0.02237\pm 0.00014 $ & $ 0.02239\pm 0.00014 $ \\ 
\textbf{\boldmath$\Omega_\mathrm{c} h^2$} & $ 0.1191\pm 0.0012  $ & $ 0.1191\pm 0.0012 $ & $ 0.1200\pm 0.0010 $ & $ 0.1199\pm 0.0010 $ \\ 
\textbf{\boldmath$w_0$} & $ -1.17^{+0.47}_{-0.63}  $ & $ -1.07\pm 0.56 $ & $ -0.74\pm 0.19 $ & $ -0.75^{+0.18}_{-0.20} $ \\ 
\textbf{\boldmath$w_a$} & $ < -0.532  $ & $ < -0.473 $ & $ -0.90^{+0.61}_{-0.51} $ & $ -0.88^{+0.61}_{-0.51} $ \\ 
\textbf{\boldmath$\tau$} & $ 0.0515\pm 0.0074 $ & $ 0.0521\pm 0.0074 $ & $ 0.0524\pm 0.0072 $ & $ 0.0527\pm 0.0073 $ \\ 
\textbf{$H_0$} & $ > 75.9  $ & $ 79\pm 10  $ & $ 66.5\pm 1.7 $ & $ 66.6\pm 1.7 $ \\ 
\textbf{$\Omega_\mathrm{m}$} & $ 0.228^{+0.022}_{-0.086} $ & $ 0.246^{+0.033}_{-0.10} $ & $ 0.324\pm 0.017 $ & $ 0.323^{+0.016}_{-0.018} $ \\ 
\textbf{$\sigma_8$}  & $ 0.924^{+0.13}_{-0.058}$ & $ 0.899^{+0.11}_{-0.083} $ & $ 0.806\pm 0.015 $ & $ 0.807\pm 0.015 $ \\ 
\textbf{$S_8$}  & $ 0.784^{+0.026}_{-0.051} $ & $ 0.793^{+0.036}_{-0.045} $ & $ 0.837\pm 0.013 $ & $ 0.836\pm 0.013 $ \\ 
\textbf{$r_\mathrm{drag}$}  & $ 147.26\pm 0.27 $ & $ 147.26\pm 0.26 $ & $ 147.10\pm 0.24 $ & $ 147.10\pm 0.24 $ \\ 
\hline
\textbf{$\chi^2$}  & $ 2771.25 $ & $ 2773.25 $ & $ 2794.06 $ & $ 2796.24 $ \\ 
\textbf{$\Delta\chi^2$}  & $ -1.21 $ & $ -2.55 $ & $ -3.48 $ & $ -3.35 $ \\ 
\textbf{$\ln \mathcal{B}_{ij}$}  & $ -0.52 $ & $ -1.09 $ & $ -3.78 $ & $ -3.49 $ \\ 
\hline\hline
\end{tabular}}
\end{center}
\caption{
Constraints at 68\% confidence level for the CPL model, obtained using different combinations of CMB, BAO, and ISW datasets. 
Parameters in bold correspond to free parameters in the analysis, while the remaining quantities are derived. 
The bottom rows report the minimum $\chi^2$, the difference $\Delta\chi^2$ relative to the $\Lambda$CDM best fit, and the logarithmic Bayes factor $\ln\mathcal{B}_{ij}$, as defined in Sec.~\ref{sec:data}.
}\label{tab:dataCPL}
\end{table*}
\begin{table*}
\begin{center}
\resizebox{0.85\textwidth}{!}{%
\begin{tabular}{c||c|c|c|c}
\hline\hline
\textbf{Parameters} & \textbf{CMB} & \textbf{CMB+ISW}&\textbf{CMB+BAO} & \textbf{CMB+BAO+ISW} \\
\hline\hline
\textbf{\boldmath$\log(10^{10} A_\mathrm{s})$}   & $ 3.038\pm 0.015 $ & $ 3.039\pm 0.014 $ & $ 3.040\pm 0.014 $ & $ 3.040\pm 0.014 $ \\ 
\textbf{\boldmath$n_\mathrm{s}$}   & $ 0.9671\pm 0.0044 $ & $ 0.9673\pm 0.0042 $ & $ 0.9646\pm 0.0037 $ & $ 0.9646\pm 0.0036 $ \\ 
\textbf{\boldmath$100\theta_\mathrm{MC}$}   & $ 1.04099\pm 0.00031 $ & $ 1.04099\pm 0.00031 $ & $ 1.04086\pm 0.00029 $ & $ 1.04087\pm 0.00029 $ \\ 
\textbf{\boldmath$\Omega_\mathrm{b} h^2$}  & $ 0.02243\pm 0.00015 $ & $ 0.02243\pm 0.00015 $ & $ 0.02235\pm 0.00013 $ & $ 0.02235\pm 0.00014 $ \\ 
\textbf{\boldmath$\Omega_\mathrm{c} h^2$}   & $ 0.1193\pm 0.0013 $ & $ 0.1192\pm 0.0012 $ & $ 0.12030\pm 0.00093 $ & $ 0.12026\pm 0.00094 $ \\ 
\textbf{\boldmath$z^{\dagger}$}   & $ > 1.98 $ & $ 2.20\pm 0.45  $ & $ > 2.58  $ & $ > 2.57 $ \\ 
\textbf{\boldmath$\tau$}   & $ 0.0525\pm 0.0077 $ & $ 0.0528\pm 0.0074 $ & $ 0.0519\pm 0.0069 $ & $ 0.0519\pm 0.0071 $ \\ 
\textbf{$H_0$}   & $ 70.65^{+0.72}_{-2.6} $ & $ 70.79^{+0.82}_{-2.6} $ & $ 68.63\pm 0.49 $ & $ 68.67\pm 0.49 $ \\ 
\textbf{$\Omega_\mathrm{m}$}   & $ 0.286^{+0.022}_{-0.0093} $ & $ 0.285^{+0.022}_{-0.010} $ & $ 0.3043\pm 0.0056 $ & $ 0.3039\pm 0.0056 $ \\ 
\textbf{$\sigma_8$}   & $ 0.8186^{+0.0060}_{-0.010} $ & $ 0.8192^{+0.0061}_{-0.0099} $ & $ 0.8154\pm 0.0058 $ & $ 0.8153\pm 0.0058 $ \\ 
\textbf{$S_8$}   & $ 0.799^{+0.026}_{-0.014} $ & $ 0.798^{+0.026}_{-0.015} $ & $ 0.8212\pm 0.0099 $ & $ 0.821\pm 0.010 $ \\ 
\textbf{$r_\mathrm{drag}$}    & $ 147.23\pm 0.27 $ & $ 147.25\pm 0.28 $ & $ 147.05\pm 0.22 $ & $ 147.05\pm 0.23 $ \\ 
\hline
\textbf{$\chi^2$}    & $ 2773.37 $ & $ 2775.48 $ & $ 2798.49 $ & $ 2800.08 $ \\ 
\textbf{$\Delta\chi^2$}    & $ 0.91 $ & $ -0.32 $ & $ 0.96 $ & $ 0.48 $ \\ 
\textbf{$\ln \mathcal{B}_{ij}$}  & $ 0.14 $ & $ 0.23 $ & $ -0.27 $ & $ -0.20 $ \\ 
\hline\hline
\end{tabular}}
\end{center}
\caption{
Constraints at 68\% confidence level for the $\Lambda_{\rm s}$CDM model, obtained using different combinations of CMB, BAO, and ISW datasets. 
Parameters in bold correspond to free parameters in the analysis, while the remaining quantities are derived. 
The bottom rows report the minimum $\chi^2$, the difference $\Delta\chi^2$ relative to the $\Lambda$CDM best fit, and the logarithmic Bayes factor $\ln\mathcal{B}_{ij}$, as defined in Sec.~\ref{sec:data}.
} \label{tab:dataLsCDM}
\end{table*}
\begin{table*}
\begin{center}
\resizebox{0.85\textwidth}{!}{%
\begin{tabular}{c||c|c|c|c}
\hline\hline
\textbf{Parameters} & \textbf{CMB} & \textbf{CMB+ISW}&\textbf{CMB+BAO} & \textbf{CMB+BAO+ISW} \\
\hline\hline
\textbf{\boldmath$\log(10^{10} A_\mathrm{s})$}   & $ 3.036\pm 0.015 $ & $ 3.038\pm 0.014 $ & $ 3.037\pm 0.014 $ & $ 3.037\pm 0.014 $ \\ 
\textbf{\boldmath$n_\mathrm{s}$}   & $ 0.9666\pm 0.0041 $ & $ 0.9671\pm 0.0043 $ & $ 0.9663\pm 0.0038 $ & $ 0.9663\pm 0.0038 $ \\  
\textbf{\boldmath$100\theta_\mathrm{MC}$}   & $ 1.04097\pm 0.00030 $ & $ 1.04100\pm 0.00031 $ & $ 1.04095\pm 0.00030 $ & $ 1.04095\pm 0.00030 $ \\ 
\textbf{\boldmath$\Omega_\mathrm{b} h^2$}  & $ 0.02243\pm 0.00015 $ & $ 0.02242\pm 0.00015 $ & $ 0.02241\pm 0.00014 $ & $ 0.02241\pm 0.00014  $ \\
\textbf{\boldmath$\Omega_\mathrm{c} h^2$}   & $ 0.1193\pm 0.0012 $ & $ 0.1192\pm 0.0012 $ & $ 0.1196\pm 0.0010 $ & $ 0.1196\pm 0.0010 $ \\
\textbf{\boldmath$a_m$}   & $ < 0.421 $ & $ < 0.579 $ & $ 0.841^{+0.026}_{-0.030} $ & $ 0.844^{+0.022}_{-0.033} $ \\ 
\textbf{\boldmath$\alpha$}   & $ < 17.0 $ & $ < 14.6 $ & $ 8.9^{+4.7}_{-2.3} $ & $ 8.9^{+4.6}_{-2.5} $ \\
\textbf{\boldmath$\beta$}   & $ < 16.2 $ & $ < 16.7 $ & $ > 16.3 $ & $ > 16.1 $ \\
\textbf{\boldmath$\tau$}    & $ 0.0512\pm 0.0076 $ & $ 0.0523\pm 0.0073 $ & $ 0.0517\pm 0.0074 $ & $ 0.0515\pm 0.0074 $ \\
\textbf{$H_0$}   & $ > 91.8 $ & $ > 89.4  $ & $ 72.7\pm 2.6 $ & $ 72.6\pm 2.6 $ \\
\textbf{$\Omega_\mathrm{m}$}   & $ 0.1728^{-0.0025}_{-0.031} $ & $ 0.1812^{-0.0015}_{-0.040} $ & $ 0.271^{+0.018}_{-0.022} $ & $ 0.272^{+0.018}_{-0.023} $ \\
\textbf{$\sigma_8$}   & $ 1.003^{+0.058}_{-0.0079} $ & $ 0.989^{+0.075}_{-0.012} $ & $ 0.855\pm 0.021 $ & $ 0.854^{+0.023}_{-0.020} $ \\
\textbf{$S_8$}   & $ 0.7524^{+0.0086}_{-0.027} $ & $ 0.757^{+0.010}_{-0.034} $ & $ 0.811\pm 0.015  $ & $ 0.812\pm 0.015 $ \\
\textbf{$r_\mathrm{drag}$}    & $ 147.22\pm 0.26 $ & $ 147.26\pm 0.28 $ & $ 147.18\pm 0.24 $ & $ 147.17\pm 0.24 $ \\
\hline
\textbf{$\chi^2$}    & $ 2770.26 $ & $ 2773.72 $ & $ 2786.21 $ & $ 2788.68 $ \\
\textbf{$\Delta\chi^2$}    & $ -2.20 $ & $ -2.09 $ & $ -11.33 $ & $ -10.92 $ \\
\textbf{$\ln \mathcal{B}_{ij}$}  & $ 0.96 $ & $ 0.51 $ & $ 4.30 $ & $ 4.57 $ \\
\hline\hline
\end{tabular}}
\end{center}
\caption{
Constraints at 68\% confidence level for the Omnipotent DE model, obtained using different combinations of CMB, BAO, and ISW datasets. 
Parameters in bold correspond to free parameters in the analysis, while the remaining quantities are derived. 
The bottom rows report the minimum $\chi^2$, the difference $\Delta\chi^2$ relative to the $\Lambda$CDM best fit, and the logarithmic Bayes factor $\ln\mathcal{B}_{ij}$, as defined in Sec.~\ref{sec:data}.
}\label{tab:dataomnipotent}
\end{table*}
\begin{figure*}[htbp]
    \centering
    \includegraphics[width=0.45\linewidth]{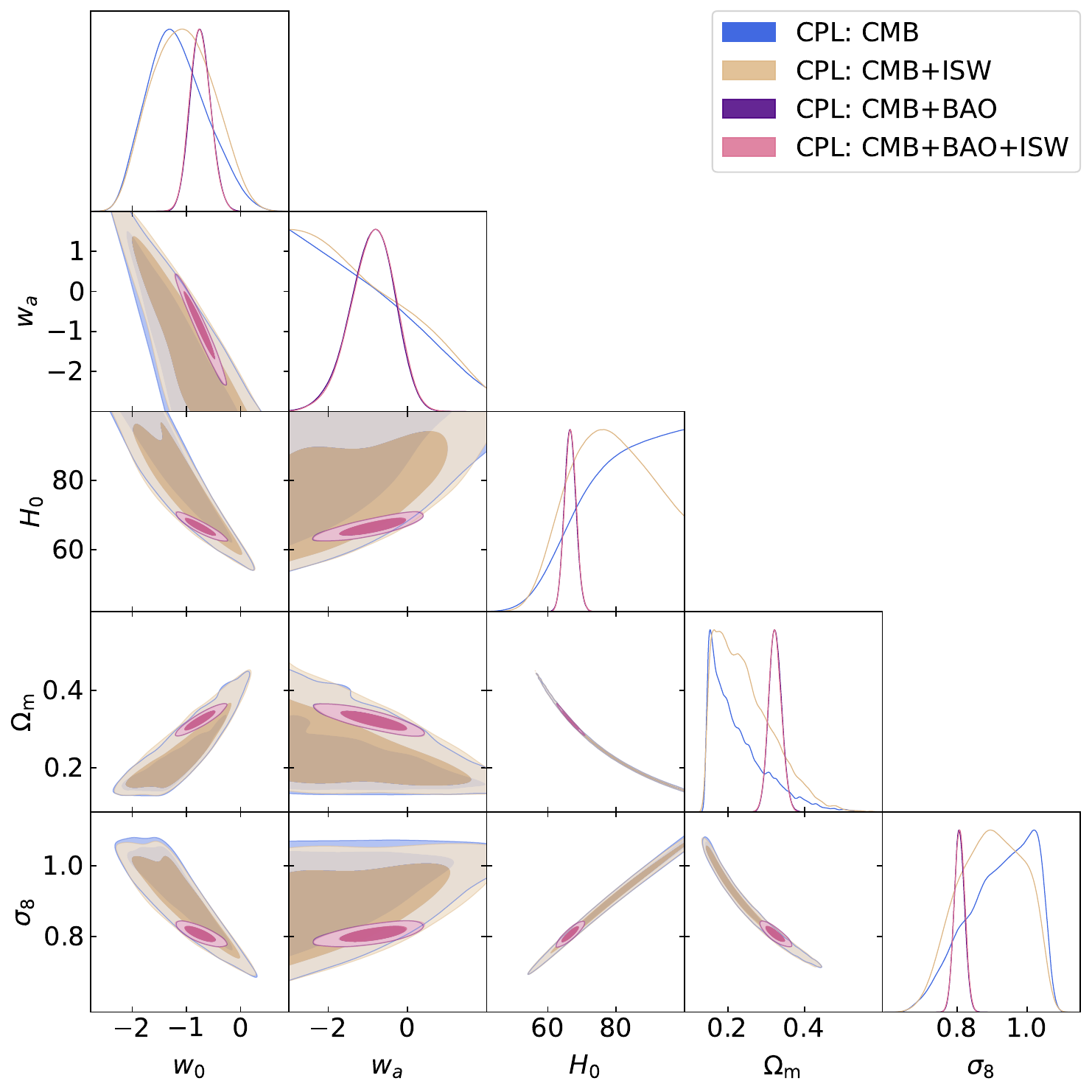}
        \includegraphics[width=0.45\linewidth]{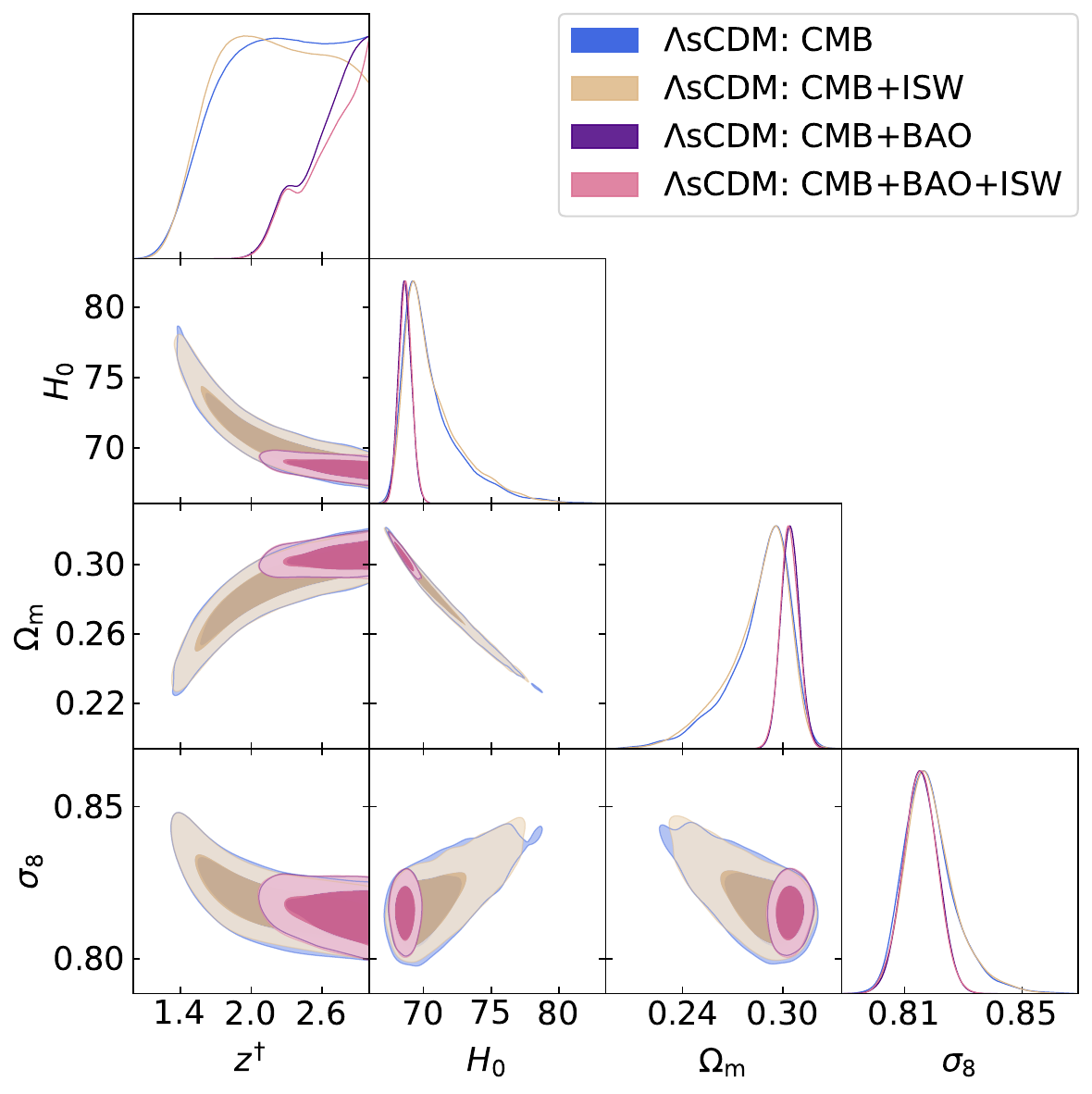}
          \includegraphics[width=0.45\linewidth]{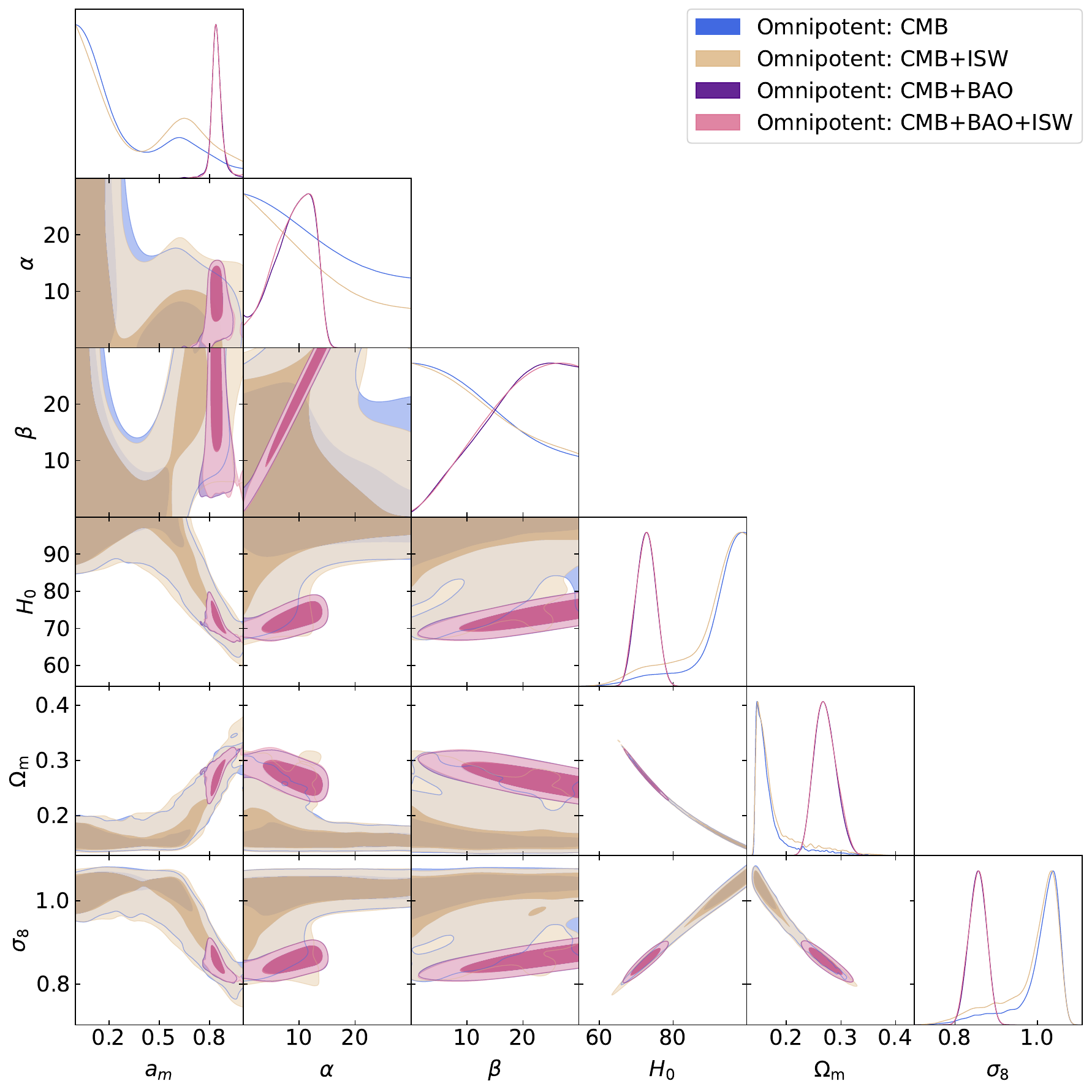}
    \caption{
Triangle plots showing the one- and two-dimensional posterior distributions for the parameters of each dark energy model, together with the derived quantities $H_0$, $\Omega_m$, and $\sigma_8$. Results are shown for the four dataset combinations: CMB, CMB+ISW, CMB+BAO, and CMB+BAO+ISW. The top left, top right, and bottom panels correspond to the CPL, $\Lambda_{\rm s}$CDM, and Omnipotent DE models, respectively.
}
    \label{fig:triangle}
\end{figure*}

\section{Data analysis results} \label{sec:results}

The main goal of this section is to assess the constraining power of the ISW data when used as a complementary probe alongside the CMB and BAO datasets. 
A full cosmological analysis of the dark energy models considered here lies beyond the scope of this section. 
For comprehensive discussions of each model and its phenomenology, we refer the reader to Refs.~\cite{Najafi:2024qzm,Specogna:2025guo,Akarsu:2023mfb,Adil:2023exv,Akarsu:2021fol}.
In this work, we focus on a set of non-standard dark energy models that have been proposed as possible solutions to the Hubble constant ($H_0$) tension by allowing for negative or sign-changing dark energy densities at certain epochs. 
Specifically, we test the $\Lambda_{\mathrm{s}}$CDM model, which features a sign-switching cosmological constant, and the Omnipotent DE model, in which the energy density can become negative in addition to crossing the phantom divide. 
For comparison, we also include the CPL parameterization, which represents a smooth, phenomenological evolution of the dark energy equation of state without negative densities.
The ISW dataset employed here corresponds to the \textit{Planck} 2022 lensing–temperature cross-correlation likelihood~\cite{Carron:2022eum}, which measures the correlation between CMB temperature anisotropies and the reconstructed lensing potential map. 
This observable provides a direct probe of the late-time decay of gravitational potentials and is particularly sensitive to the dynamics of dark energy at low redshifts. 
Unlike the galaxy cross-correlation approach, which depends on external large-scale structure tracers, this internal CMB measurement relies solely on \textit{Planck} data, ensuring homogeneous coverage and well-characterized systematics. 
Although the statistical weight of the ISW–lensing correlation is modest compared to that of the primary CMB anisotropies, its inclusion offers an independent and complementary constraint on models that modify the late-time expansion history.

We begin our analysis with the CPL parameterization. Table~\ref{tab:dataCPL} summarizes the constraints on the cosmological parameters obtained using different combinations of CMB, BAO, and ISW datasets at the 68\% confidence level (CL). 
Focusing first on the CMB-only case, we find that only a lower limit on the Hubble constant is obtained, $H_0 > 75.9~\mathrm{km\,s^{-1}\,Mpc^{-1}}$ (68\% CL), as expected given the highly phantom nature of the best-fit value of $w_0$ (see, e.g., Ref.~\cite{Najafi:2024qzm}). 
Due to the well-known degeneracy between $H_0$ and $\Omega_m$ (see also Fig.~\ref{fig:triangle}), this high $H_0$ value corresponds to a low matter density, $\Omega_m = 0.228^{+0.022}_{-0.086}$ (68\% CL). 
The CMB-only data also place only an upper limit on the parameter $w_a$, leaving the time variation of the dark energy equation of state largely unconstrained.
When the ISW data are added to the CMB likelihood, the main cosmological parameters remain consistent. 
However, the inclusion of the ISW information allows $H_0$ to be properly constrained, yielding $H_0 = 79 \pm 10~\mathrm{km\,s^{-1}\,Mpc^{-1}}$, whereas CMB data alone could only provide a lower limit (see top left panel of Fig.~\ref{fig:triangle}). The uncertainties on all cosmological and dark energy parameters remain essentially unchanged when the ISW likelihood is added. The goodness of fit improves modestly, with the relative $\Delta\chi^2$ difference with respect to $\Lambda$CDM changing from $\Delta\chi^2 = -1.21$ (CMB-only) to $\Delta\chi^2 = -2.55$ (CMB+ISW), while the Bayesian evidence remains inconclusive.
The addition of BAO data significantly improves the overall constraining power, breaking the degeneracies present in the CMB-only case. 
In the CMB+BAO combination, the dark energy parameters shift toward a quintessence-like regime with $w_0 = -0.74 \pm 0.19$ and $w_a = -0.90^{+0.61}_{-0.51}$ at 68\% CL, while $H_0$ decreases to $66.5 \pm 1.7~\mathrm{km\,s^{-1}\,Mpc^{-1}}$, in $3.6\sigma$ tension with the H0DN consensus distance-ladder measurement of $H_0 = 73.50 \pm 0.81~\mathrm{km\,s^{-1}\,Mpc^{-1}}$~\cite{H0DN:2025lyy}. 
Including ISW data in this combination (CMB+BAO+ISW) produces nearly identical results, both in parameter estimates and in the relative $\chi^2$ improvement, confirming that once BAO constraints are included, the ISW information adds little additional statistical weight. 
Overall, the CMB+BAO(+ISW) combinations yield higher $\Omega_m$ values and tightly constrain the dark energy evolution, with results that are remarkably stable across dataset choices.

We now focus on the sign-switching case, $\Lambda_{\rm s}$CDM. Table~\ref{tab:dataLsCDM} presents the constraints obtained using CMB, CMB+ISW, CMB+BAO, and CMB+BAO+ISW data combinations at the 68\% CL. 
Given the sign-switching nature of $\Lambda_{\rm s}$CDM, a late-time transition of the cosmological constant’s sign is expected to produce higher values of $H_0$ and correspondingly lower $\Omega_m$. 
As already discussed in Ref.~\cite{Akarsu:2023mfb}, CMB data alone are unable to place upper bounds on $z^\dagger$, and we recover consistent behavior here, with the marginalized one-dimensional posterior exhibiting a plateau at higher values, as seen in Fig.~\ref{fig:triangle}.
In this case, we obtain $H_0 = 70.65^{+0.72}_{-2.6}~\mathrm{km\,s^{-1}\,Mpc^{-1}}$, which alleviates the $H_0$ tension, together with $\Omega_m = 0.286^{+0.022}_{-0.0093}$ (68\% CL). 
However, the minimum $\chi^2$ value for this dataset combination is slightly worse than for $\Lambda$CDM, and the Bayesian evidence does not show any significant preference between the two models.
When the ISW data are added to the CMB likelihood, the posteriors are barely effected as seen in Fig.~\ref{fig:triangle}.
The minimum $\chi^2$ improves marginally with the inclusion of ISW data, while the Bayesian evidence remains statistically inconclusive. 
The mean values of the remaining parameters remain stable within uncertainties.
When BAO data are added, the situation changes significantly. 
For both CMB+BAO and CMB+BAO+ISW combinations, the lower limit on $z^\dagger$ is considerably higher, pushing the negative cosmological constant deeper in the matter dominated era and forcing it to exhibit a phenomenology similar to a $\Lambda$CDM-like behaviour. 
This demonstrates that the BAO data, which strongly constrain the background expansion, drive the analysis toward the $\Lambda$CDM limit. 
As a consequence, the minimum $\chi^2$ now favours $\Lambda$CDM despite the extra degree of freedom in $\Lambda_{\rm s}$CDM (this is possible since the models are not nested). 
The Hubble constant decreases to $H_0 = 68.63\pm0.49~\mathrm{km\,s^{-1}\,Mpc^{-1}}$ ($68.67\pm0.49$ when ISW data are included), restoring the $H_0$ tension to about the $5\sigma$ level. 
In the combination CMB+BAO, the constraints are already dominated by the geometric information from BAO, which sharply restricts the allowed parameter space. Also in this case, adding ISW data leaves the results effectively unchanged, demonstrating that BAO data already saturate the constraining power, leaving the ISW likelihood with only a minimal impact (see also top right panel of Fig.~\ref{fig:triangle}).

Before discussing the results for the Omnipotent DE model, we encourage the reader to refer to Ref.~\cite{Specogna:2025guo} for an updated analysis and discussion of prior effects in this phenomenology, as well as to the earlier work~\cite{DiValentino:2020naf} for details on the parameter bimodality and its physical interpretation. 
Table~\ref{tab:dataomnipotent} summarizes the constraints on the cosmological parameters obtained using CMB, CMB+ISW, CMB+BAO, and CMB+BAO+ISW combinations at the 68\% CL.
In the case of CMB-only, the Omnipotent DE parameters remain largely unconstrained, with only upper limits on $a_m$, $\alpha$, and $\beta$. 
The upper limit on $a_m$ arises from the known bimodality of this parameter, as we can see in the bottom panel of Fig.~\ref{fig:triangle}, which cannot be resolved with CMB data alone or even with the addition of lensing, as already shown in Ref.~\cite{DiValentino:2020naf}. 
In this case, while the fit slightly improves compared to $\Lambda$CDM, the gain in $\chi^2$ is insufficient to overcome the Occam’s razor penalty associated with the additional degrees of freedom, and the Bayesian evidence still disfavors the model.
When ISW data are added to the CMB likelihood, the overall shape of the posterior distributions becomes better defined.  
The uncertainties on most parameters are reduced, and the upper limit on $a_m$ is slightly relaxed, as is the lower limit on $H_0$. 
Nevertheless, $a_m$ remains bimodal, with the second peak slightly increased (see in the bottom panel of Fig.~\ref{fig:triangle}), and the inclusion of ISW data, while marginally improving $\chi^2_{\rm min}$, does not lead to any significant change in the Bayesian evidence, which continues to show no statistical preference for Omnipotent DE over $\Lambda$CDM. 
These results suggest that the ISW information helps reduce parameter degeneracies slightly but is insufficient to break them to the extent of providing new insights within the extended parameter space of this model.
When BAO data are included, the situation improves considerably. 
The second peak in the posterior of $a_m$ is selected, leading to a well-defined constraint on the transition scale factor, $a_m = 0.841^{+0.026}_{-0.030}$ ($0.844^{+0.022}_{-0.033}$ with ISW), corresponding to a late-time phantom crossing. 
The parameter $\alpha$ is now constrained, while $\beta$ remains poorly determined, with only a lower limit. 
In this configuration, the Hubble constant is tightly constrained to $H_0 = 72.7\pm2.6$~$\mathrm{km\,s^{-1}\,Mpc^{-1}}$, in excellent agreement with the H0DN determination, thereby resolving the $H_0$ tension within $1\sigma$. 
The improvement in the best-fit $\chi^2$ is significant and points toward a better fit to the data compared to $\Lambda$CDM. 
However, when penalizing for the additional parameters, the Bayesian evidence still does not favour Omnipotent DE, highlighting once again that statistical model comparison disfavors complexity despite a better fit. 
Finally, the inclusion of ISW data in the BAO combination produces virtually no change in either the central values or the error bars, confirming that, in this high-dimensional parameter space, the ISW–lensing cross-correlation adds little additional constraining power. 

In summary, the comparison of $\Delta \chi^2$ and Bayesian evidence values across all data combinations indicates that the inclusion of ISW information provides, at best, marginal improvements in the overall fit to the data. The Omnipotent DE model achieves the largest $\chi^2$ reduction, as expected given its greater number of degrees of freedom, but this improvement is not sufficient to overcome the statistical penalty associated with model complexity. In contrast, the Bayesian evidence remains generally inconclusive, with no model exhibiting a decisive preference over $\Lambda$CDM across the tested datasets. For all models, and especially within the extended Omnipotent parameter space, the ISW data modestly refine uncertainties but do not add significant constraining power or shift central values.  
Overall, while the ISW–lensing cross-correlation serves as a valuable late-time probe sensitive to the decay of gravitational potentials, its statistical weight remains limited compared to CMB and BAO datasets in constraining current dark energy models. Future high–signal-to-noise ISW measurements, particularly from upcoming CMB and LSS cross-correlation surveys, will be essential to fully exploit the discriminating potential of this observable.

\section{Conclusions}
\label{sec:conclusion}

In this work, we have carried out a multi–probe ISW analysis of three dark energy scenarios beyond the standard $\Lambda$CDM model: the CPL parametrization, the $\Lambda_{\rm s}$CDM model with a sign-switching cosmological constant, and the Omnipotent DE phenomenology, which allows for negative dark energy densities and phantom crossings. Our goal was to investigate how these models modify late-time cosmological observables that are sensitive to the evolution of the Weyl potential, and to assess the constraining power of current ISW data when combined with CMB and BAO measurements.

At the level of primary CMB anisotropies, all three models reproduce the acoustic peak structure of the temperature power spectrum and remain effectively indistinguishable from $\Lambda$CDM once cosmic variance at low multipoles is taken into account. Differences become more apparent when considering the matter power spectrum and secondary anisotropies. Using CMB-only best-fit parameters, the various dark energy models predict noticeably different matter clustering amplitudes, reflecting degeneracies between $H_0$, $\Omega_m$, and the dark energy sector. The addition of BAO data efficiently breaks these degeneracies, driving $\Lambda_{\mathrm{s}}$CDM close to $\Lambda$CDM at large scales ($k \lesssim 0.01$), while CPL remains systematically below $\Lambda$CDM for all $k$, and Omnipotent DE stays above it across the entire range, especially at $k \lesssim 0.01$.

We then studied late-time probes of the gravitational potential through the ISW effect and the lensing–ISW bispectrum. For the ISW–galaxy cross-correlation, we showed that an SDSS-like survey, which peaks at low redshift, is more sensitive to model-dependent modifications of the potential decay than a Euclid-like survey, whose higher-redshift kernel overlaps less with the ISW window. Dynamical dark energy models generally alter the amplitude and scale dependence of $C_\ell^{Tg}$, but BAO-calibrated best fits tend to reduce these differences. In the Omnipotent DE case, within the prior volume of the parameter space, the parameter $a_m$ (the phantom-crossing scale factor) plays the dominant role in shifting and suppressing the ISW contribution at low multipoles, while $\alpha$ and $\beta$ introduce more subtle changes in amplitude and scale dependence. In $\Lambda_{\rm s}$CDM, increasing the transition redshift $z^{\dagger}$ gradually brings the ISW–related spectra closer to the $\Lambda$CDM prediction.

The lensing–ISW bispectrum provides a complementary, higher-order probe of late-time physics. We computed the reduced bispectrum $b_{\ell_1\ell_2\ell_3}$ for all three models and found that, even though their CMB power spectra are nearly indistinguishable at high multipoles, and their differences at low multipoles lie within the cosmic-variance-limited region, the bispectrum nonetheless exhibits clear, model-dependent oscillatory patterns with different amplitudes at the peaks and troughs. In particular, $\Lambda_{\rm s}$CDM and Omnipotent DE can generate stronger oscillations than CPL and $\Lambda$CDM in some configurations, reflecting a more pronounced coupling between long-wavelength ISW modes and small-scale lensing. Parameter scans showed that, within the prior volume of the parameter space, $a_m$ in Omnipotent DE has the largest impact on the bispectrum scale and amplitude, with $\alpha$ and $\beta$ acting mainly as global enhancers or suppressors, while variations in $z^{\dagger}$ in $\Lambda_{\rm s}$CDM produce non-trivial but controlled shifts relative to the $\Lambda$CDM shape.

Finally, using real ISW data from the \textit{Planck} lensing–temperature cross-correlation likelihood in combination with CMB and BAO, we constrained the parameter spaces of the three dark energy models. For CPL, ISW data constrain $H_0$ better ($79 \pm 10~\mathrm{km\,s^{-1}\,Mpc^{-1}}$, see also Fig.~\ref{fig:triangle}) compared to the anisotropy only data, but slightly relax the constraints on $w_0$ and $w_a$, without yielding a significant improvement in $\chi^2$ or Bayesian evidence. However, the inclusion of BAO drives the model back to $H_0 \simeq 66.6~\mathrm{km\,s^{-1}\,Mpc^{-1}}$. For $\Lambda_{\rm s}$CDM, inclusion of the ISW data has negligible impact on the parameter posteriors with or without the BAO data, and the BAO data pushes the model behavior towards $\Lambda$CDM preventing it from significantly alleviating the $H_0$ tension. The Omnipotent DE model retains large parameter degeneracies, but when BAO are included, the phantom-crossing scale factor becomes well constrained ($a_m \simeq 0.84$) and $H_0 \simeq 72$–$73~\mathrm{km\,s^{-1}\,Mpc^{-1}}$, consistent with local values, albeit still disfavored by Occam’s razor and Bayesian evidence. Overall, while ISW data may modestly refine some parameters, they do not yet add decisive constraining power beyond CMB and BAO.

Taken together, our results highlight both the promise and the current limitations of ISW-based probes for testing exotic dark energy scenarios with negative or sign-switching energy densities. On the one hand, two- and three-point CMB statistics that are sensitive to the late-time Weyl potential, such as $C_\ell^{Tg}$ and the lensing–ISW bispectrum, encode distinctive signatures of these models, even when their background expansion is nearly indistinguishable from $\Lambda$CDM. On the other hand, with present data, the statistical impact of the ISW–lensing cross-correlation is modest compared to CMB and BAO, and no strong Bayesian preference for any of the extended models is found. Upcoming CMB and large-scale structure surveys, if they are to enhance the precision of ISW and lensing adequately on top of improved control of systematics, could be crucial to decisively establish whether dynamical dark energy scenarios that may even exhibit negative and/or oscillatory densities are supported or ruled out by observations.

\textbf{Data Availability}: Data underlying this research will be available upon reasonable request after the publication of this article.

\section{Acknowledgements}
The authors acknowledge insightful discussions with Ali Tizfahm. EDV is supported by a Royal Society Dorothy Hodgkin Research Fellowship.
We acknowledge the IT Services at The University of Sheffield for the provision of services for High Performance Computing.
This article is based upon work from the COST Action CA21136 - ``Addressing observational tensions in cosmology with systematics and fundamental physics (CosmoVerse)'', supported by COST - ``European Cooperation in Science and Technology''.

\

\bibliography{sample631,biblio}{}
\bibliographystyle{unsrt}

\end{document}